\documentclass[pra,reprint,amsmath,amssymb,aps,superscriptaddress]{revtex4-2}
\usepackage{graphicx}
\usepackage{hyperref}
\usepackage{color}
\usepackage{makecell}
\usepackage{listings}
\begin{document}
\title{A $\boldsymbol{2d \times d \times d}$ Spacetime Volume Implementation \\of a Logical S Gate in the Surface Code}

\author{Yuga Hirai}
\email{yuga917@keio.jp}
\affiliation{Department of Applied Physics and Physico-Informatics, Keio University, Kanagawa, Japan}

\author{Shota Ikari}
\email{shota-ikari@g.ecc.u-tokyo.ac.jp}
\affiliation{Graduate School of Information Science and Technology, The University of Tokyo, Tokyo, Japan}

\author{Yosuke Ueno}
\email{yosuke.ueno@riken.jp}
\affiliation{RIKEN Center for Quantum Computing, Saitama, Japan}
\affiliation{Graduate School of Information Science and Technology, The University of Tokyo, Tokyo, Japan}

\author{Yasunari Suzuki}
\email{yasunari.suzuki@riken.jp}
\affiliation{RIKEN Center for Quantum Computing, Saitama, Japan}

\date{\today}

\begin{abstract}
The logical S gate implemented via twist defect braiding in the surface code is one of the major sources of overhead in fault-tolerant quantum computing, since an S-gate correction is required in every logical T-gate teleportation. 
Existing logical S-gate implementations require spacetime volumes of \(2d \times 2d \times d\) or \(2d \times 1.5d \times d\), where $d$ is the code distance of the surface code.
To the best of our knowledge, their circuit-level implementations have not yet been shown, hindering quantitative comparisons of fault distances and logical error rates. 
In this work, we provide these missing circuit-level implementations. 
Additionally, we propose a novel twist defect braiding protocol that reduces the spacetime volume to \(2d \times d \times d\). 
First, we construct an implementation of the proposed method using constant-length non-local gates, and then refine it to utilize only nearest-neighbor two-qubit gates on a square grid, without requiring additional two-qubit gate depth beyond that of standard syndrome extraction circuits. 
Through numerical simulations, we evaluate the fault distances and logical error rates for both existing and proposed methods. 
Our results show that, although the proposed method reduces the fault distance by one or three, its logical error rates remain comparable to those of existing methods at large code distances (\(d \ge 5\)) and at physical error rates near \(p = 10^{-3}\). 
This demonstrates that the proposed method is promising for near-term fault-tolerant quantum computing.

\end{abstract}

\maketitle

\section{Introduction}
In recent years, quantum computing devices have demonstrated fault-tolerant logical operations using several quantum error correction~(QEC) codes~\cite{lacroix2025scaling,google2025quantum,rosenfeld2025magic,zhang2025demonstrating}. 
However, reducing the spacetime overhead of logical operations and improving their fidelity remain essential challenges for achieving quantum advantage. 
Among QEC codes, the surface code is considered the most promising because it requires only spatially local interactions for syndrome extraction, has a high error-correction threshold~\cite{stephens2014fault,fowler2009high}, and admits efficient decoding~\cite{Higgott2025sparseblossom}. 
The surface code also enables the implementation of logical operations using only local interactions via a technique known as code deformation~\cite{bombin2009quantum}, in which the logical operator is transformed while the code is deformed into another code and then restored to its original form. 
To date, numerous studies have investigated the implementation of logical H, S, CNOT, and T gates~\cite{geher2024error,brown2017poking,sahay2025error,yoder2017surface,moussa2016transversal,chen2024transversal,gidney2024inplace,bombin2023logical,horsman2012surface,gidney2024magic,hirano2025efficient,vaknin2025efficient}, which are required for universal quantum computation.

This paper focuses on implementing logical S gates on surface codes using nearest-neighbor interactions or constant-length non-local interactions in a square grid. 
Such logical S gates can be implemented by braiding twist defects~\cite{brown2017poking,yoder2017surface,litinski2018lattice,chamberland2022circuit}, which we refer to as \textit{twist-defect braiding}. 
Some effort has been devoted to reducing the spacetime volume of logical S gates based on twist-defect braiding. 
Bombín {\it et al.} achieved a spacetime volume of $2d\times 2d\times d$~(Fig.~\ref{fig:defect_diagram_bombin})~\cite{bombin2023logical}, where $d$ is the code distance. 
Gidney {\it et al.} then reduced this to $2d\times 1.5d\times d$ through a careful analysis of Y-type error asymmetry~(Fig.~\ref{fig:defect_diagram_gidney})~\cite{gidney2024inplace}. 
However, the logical S gate remains a major source of overhead in fault-tolerant quantum computation, since an S-gate correction is required in every T gate teleportation. 
This overhead becomes even more pronounced as T gates become less expensive through magic-state cultivation~\cite{gidney2024magic}. 
Therefore, further reducing the spacetime volume of logical S gates in surface codes is desirable.

Another issue in developing logical S-gate implementations is that, to the best of our knowledge, no circuit-level implementation~(e.g., Stim circuits~\cite{gidney2021stim}) exists publicly. 
This makes it difficult to quantitatively compare the properties of different methods, such as logical error rates, fault distances (i.e., the effective code distance under circuit-level noise), and the consistency of the cycle count required for synchronization with other surface-code operations~\cite{maurya2025synchronization}.

In this work, we address both of these issues. 
First, we present a method to reduce the spacetime volume of logical S gates based on twist-defect braiding to $2d \times d \times d$. 
This implementation achieves fault distance $d-1$ when constant-length non-local interactions are available. 
Even if the available physical operations are restricted to nearest-neighbor interactions, it still achieves a fault distance of $d-3$ using CXSWAP gates, without requiring additional two-qubit gate depth per syndrome-extraction cycle. 
Our improvement is based on the analysis of the Y-type error asymmetry shown in Ref.~\cite{gidney2024inplace}. 
Although twist defects are geometrically close in the defect diagrams, the fault distance remains $d \pm O(1)$ because error chains cannot form a short logical error. 
Next, we provide explicit implementations of both our proposals and two existing methods mentioned earlier, enabling comparisons in terms of fault distance, logical error rate, and per-cycle gate depth. 
Our Stim circuits are publicly available on GitHub~\cite{stimcircuit_github}.

We numerically evaluate the performance of our proposals and compare it with existing methods. 
Although our constructions slightly reduce the fault distance, the resulting logical error rates are comparable to those of existing methods when the code distance is sufficiently large ($d \geq 5$) and the physical error rate is around $10^{-3}$. 
We attribute this to the fact that the number of logical error chains shorter than $d$ is small and therefore contributes little to the total logical error rate. 
This indicates that the proposed methods can serve as an efficient implementation of logical S gates with a practically negligible drawback. 
We therefore believe that it represents one of the most highly efficient and practical approaches for implementing logical S gates in the surface code.

Our contributions are summarized as follows:
\begin{itemize}
    \item[(I)] We propose a method that implements the logical S gate with a spacetime volume of $2d\times d\times d$, keeping the reduction of the fault distance constant against the code distance and avoiding synchronization issues.
    \item[(II)] We introduce a technique to implement the proposed method using only local gates.
    \item[(III)] We provide Stim circuit implementations for all methods, including existing ones, and evaluate their fault distance upper bounds and logical error rates under circuit-level noise.
\end{itemize}
The remainder of this paper is organized as follows. 
In Sec.~\ref{sec:preliminaries}, we review the preliminaries necessary for the proposed method, including the surface code, code deformation, decoding graph, detector diagram, and the defect diagram. 
In Sec.~\ref{sec:motivation}, we describe the existing methods for the logical S gate and discuss their drawbacks. 
In Sec.~\ref{sec:proposed}, we present the proposed method and its local-gate implementation in detail. 
In Sec.~\ref{sec:results}, we present numerical results comparing fault distances and logical error rates across all methods under circuit-level noise. Finally, in Sec.~\ref{sec:conclusion}, we summarize our contributions and discuss future directions.

\section{Preliminaries}\label{sec:preliminaries}
In this section, we provide the preliminaries necessary for understanding the implementations of the logical S gate in the surface code. 
Since the preliminary information largely overlaps with our previous work~\cite{hirai2026no}, please refer to the preliminaries for more detailed explanations.

\subsection{Surface Code and Detector Diagrams}
QEC is necessary for suppressing logical error rates to a sufficiently small value, and stabilizer codes are the most common framework for constructing QEC codes. 
In stabilizer codes, the code space is defined as the simultaneous $+1$ eigenspace of a set of commuting Pauli operators, which are called \textit{stabilizers}. 
Errors occurring during computation are detected and estimated by repeatedly measuring the eigenvalues of these stabilizers.

This paper focuses on a specific variant of the surface code known as the rotated surface code. 
Rotated surface codes are considered one of the most promising QEC codes because they require only nearest-neighbor interactions for syndrome extraction, have a high error-correction threshold, and admit efficient decoding algorithms~\cite{higgott2022pymatching}. 
Consequently, they have been the subject of several recent experimental demonstrations~\cite{google2025quantum}.

\begin{figure}[t]
    \centering
    \includegraphics[scale=0.28]{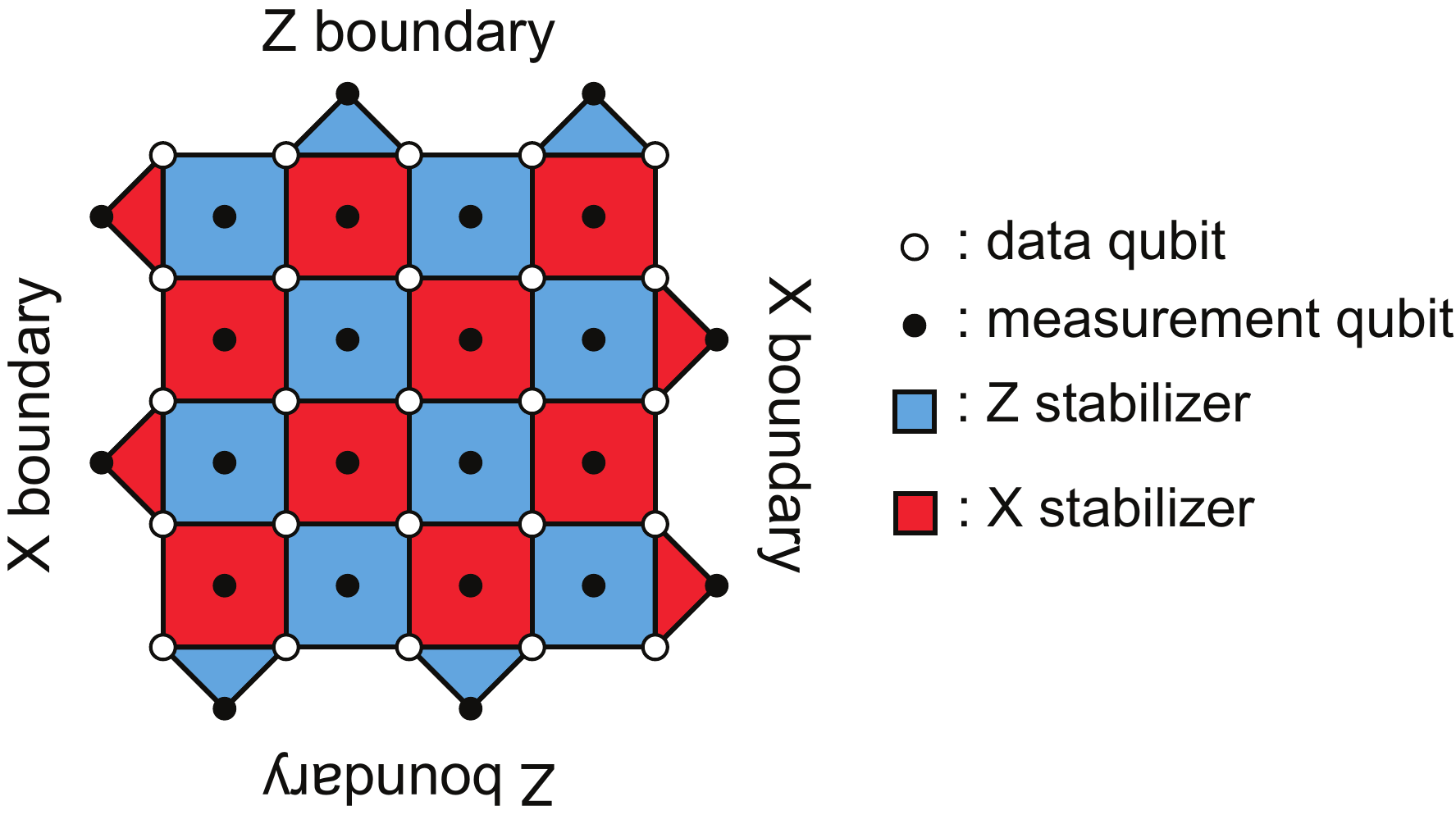}
    \caption{Rotated surface code.}
    \label{fig:surface_code}
\end{figure}

Figure~\ref{fig:surface_code} shows a rotated surface code. 
The white and black circles represent data and measurement qubits, respectively. 
The blue (red) shapes represent the Z (X) stabilizers. 
The boundaries terminated by the Z~(X) stabilizers are called the Z~(X) boundaries. 
The logical Z~(X) operator is defined as a chain connecting the two disjoint Z~(X) boundaries. 
To keep the code distance $d$, i.e., detect any non-trivial Pauli error acting on less than $d$ qubits, distinct pairs of boundaries with the same color must be separated by $d$ distance. 
We refer to the square surface code shown in Fig.~\ref{fig:surface_code} as a \textit{memory configuration}, as its primary role is to preserve logical information.

\begin{figure}[t]
    \centering
    \includegraphics[scale=0.15]{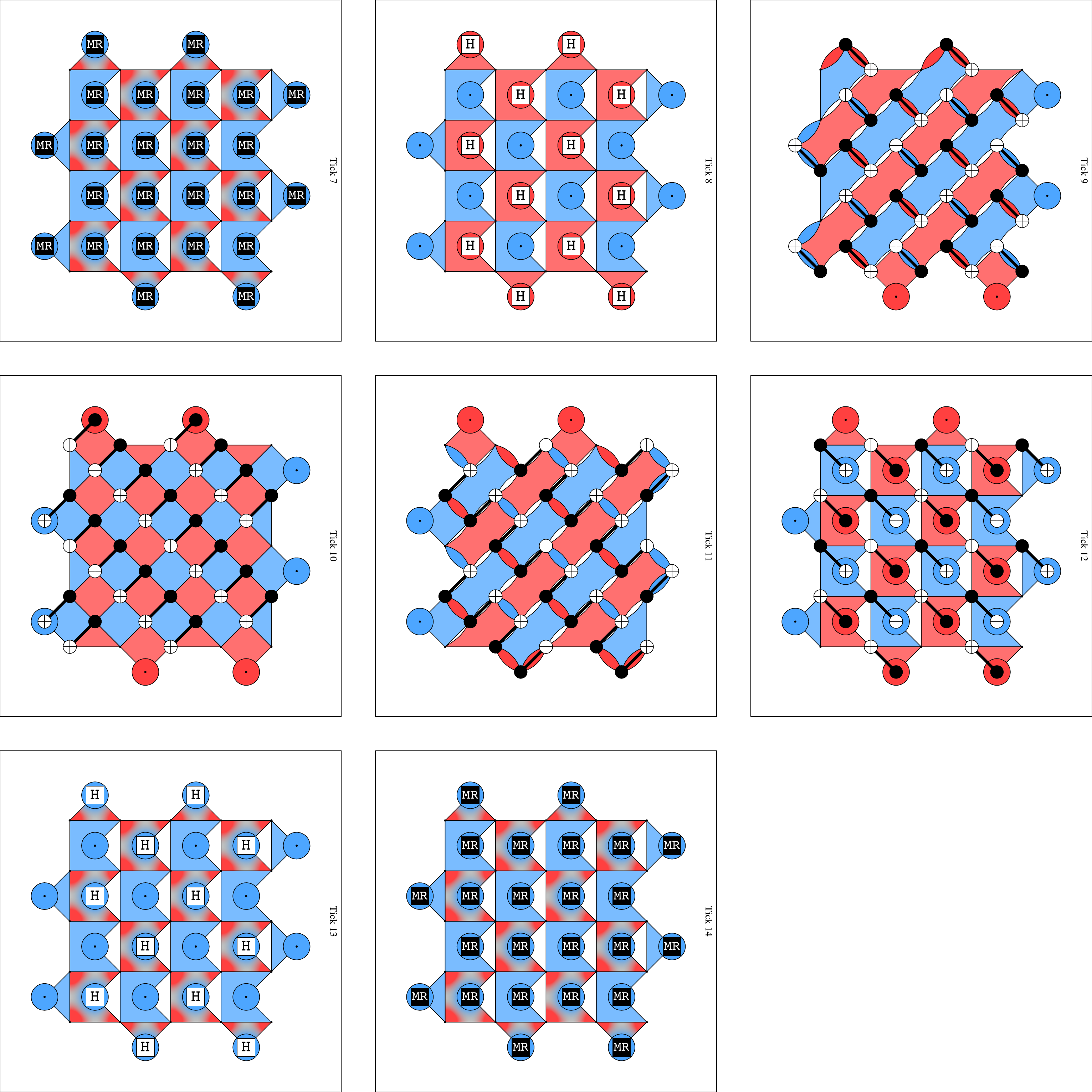}
    \caption{Detector diagram of the surface code.}
    \label{fig:detector_diagram}
\end{figure}

We need to implement multi-body stabilizer Pauli measurements as a sequence of two-qubit gates and single-qubit measurements. 
This implementation is non-trivial, as a single-qubit Pauli error occurring during these multi-body Pauli measurements may result in multi-qubit errors due to error propagation. 
Such propagated errors, known as \textit{hook errors}, can reduce the effective code distance, referred to as the \textit{fault distance}.
A \textit{detector diagram} is a useful framework for explicitly representing implementations with physical gates and monitoring the fault distance~\cite{gidney2021stim,mcewen2023relaxing}. 
In this framework, \textit{detectors} are defined as the XOR values of raw Pauli measurement outcomes that become zero in the absence of Pauli errors. 
A detector diagram visualizes each detector as a region that can identify Pauli errors acting on the target qubits. 
An example of a detector diagram for the surface code is shown in Fig.~\ref{fig:detector_diagram}. 
The blue and red shapes represent the Z- and X-type detectors, respectively. 
An error occurring in the circuit flips the detectors of the anti-commuting Pauli type.

\subsection{Code deformation and Defect Diagrams}
{\it Code deformation} is a framework for implementing logical Clifford gates on the surface code using only local gates (i.e., the nearest-neighbor interactions). 
This is particularly vital in two-dimensional (2D) qubit-array architectures, where transversal implementation of most logical Clifford operations is not possible. 
Code deformation is a technique that temporarily transforms the code into another code, thereby transforming the logical operators, and then restores the original code while preserving the transformed logical operators.

\begin{figure}[h]
    \centering
    \includegraphics[scale=0.27]{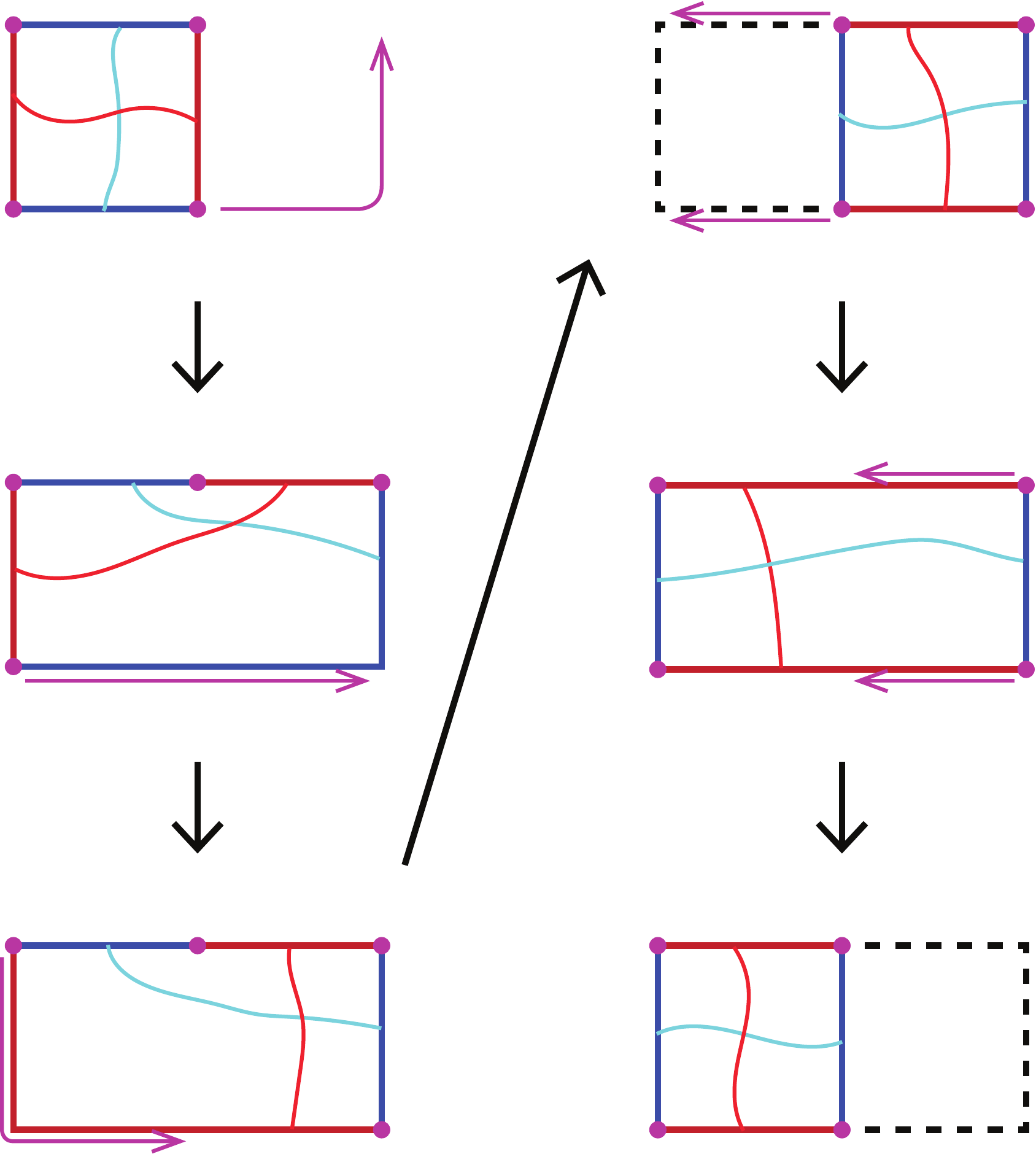}
    \caption{2D defect diagram of patch rotation. }
    \label{fig:patch_rotation_schematic}
\end{figure}

The process of code deformation can be represented with defect diagrams. 
An example of code deformation represented via this diagram is shown in Fig.~\ref{fig:patch_rotation_schematic}. 
In this figure, the dark blue (dark red) lines represent the Z (X) boundaries of the surface code. 
The purple circles represent the twist defects at which Y-type error chains terminate. 
The light blue (light red) line represents the logical Z (X) operator. 
The purple arrows represent the movement of the twist defects at each step. 
This process eventually rotates the surface code to exchange the vertical and horizontal boundaries.
This operation, known as patch rotation, is essential for implementing the logical Hadamard (H) gate~\cite{geher2024error}. 
To maintain the fault distance, an additional $d \pm O(1)$ rounds of syndrome extraction are required in some steps, but these are omitted here for simplicity.

\begin{figure}[t]
    \centering
    \includegraphics[scale=0.13]{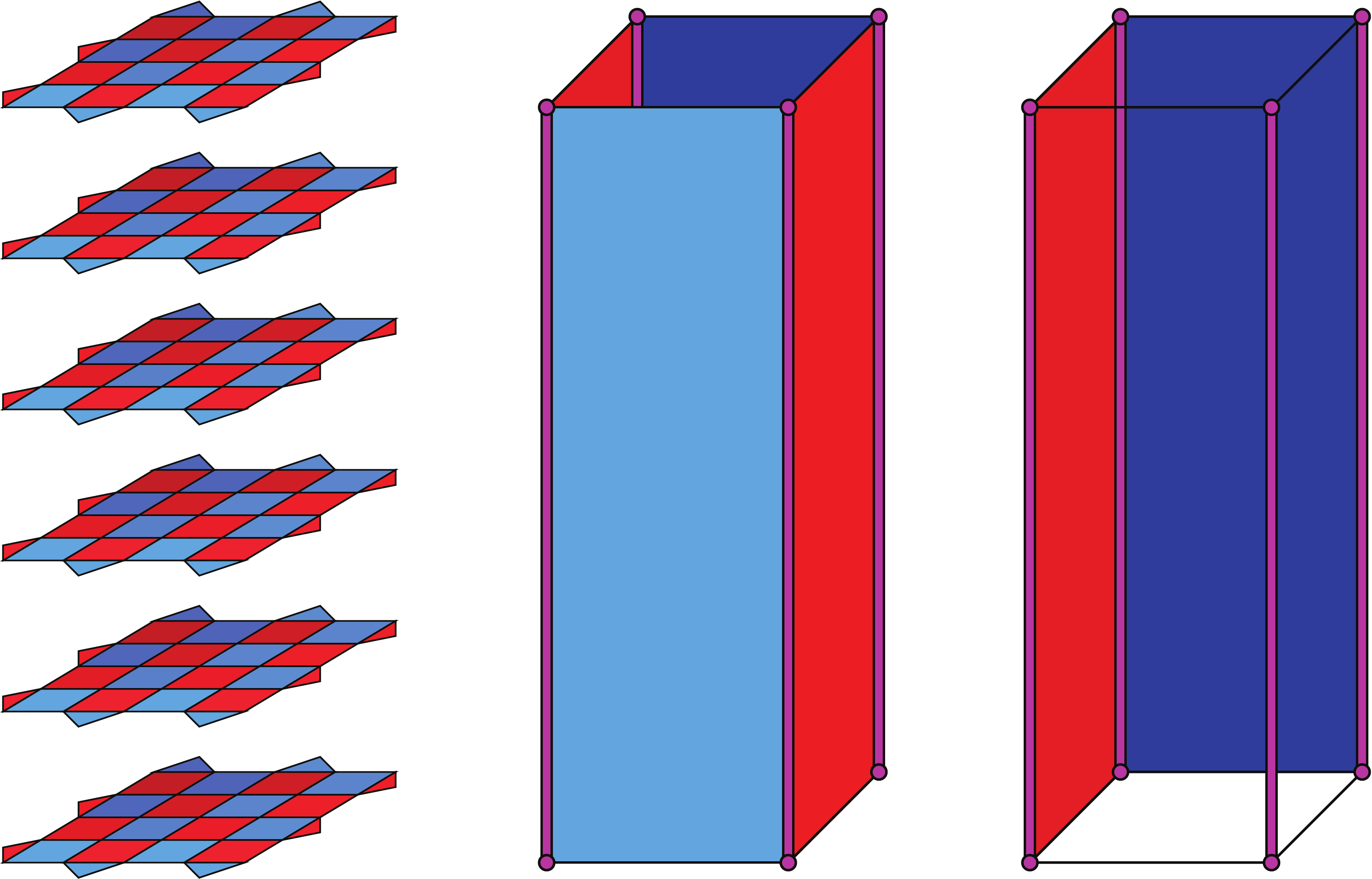}
    \caption{3D defect diagram of the memory configuration. The time goes upward.}
    \label{fig:defect_diagram}
\end{figure}

Alternatively, code deformation procedures can be visualized in 3D spacetime rather than as a sequence of 2D snapshots. 
For example, a defect diagram of the memory configuration in three dimensions is shown in Fig.~\ref{fig:defect_diagram}. 
The left side of the figure depicts the original 2D surface code stacked along the time direction. 
The middle one shows the corresponding topological 3D spacetime diagram. The right one is identical to the middle one, but with the front walls rendered transparent. 

The blue (red) walls represent the Z (X) boundaries of the surface code, and the purple lines indicate twist defects. 
In this representation, Z (X) errors terminate at the blue (red) walls, while Y errors terminate at the purple lines. 
Thus, the minimum distance between disjoint walls of the same color and twist pairs is equal to the code distance $d$ throughout this process. 
However, it should be emphasized that, in the presence of hook errors, simply maintaining a spatial separation of $d$ is not sufficient to guarantee a fault distance of $d$. 

\begin{figure}[t]
    \centering
    \includegraphics[scale=0.15]{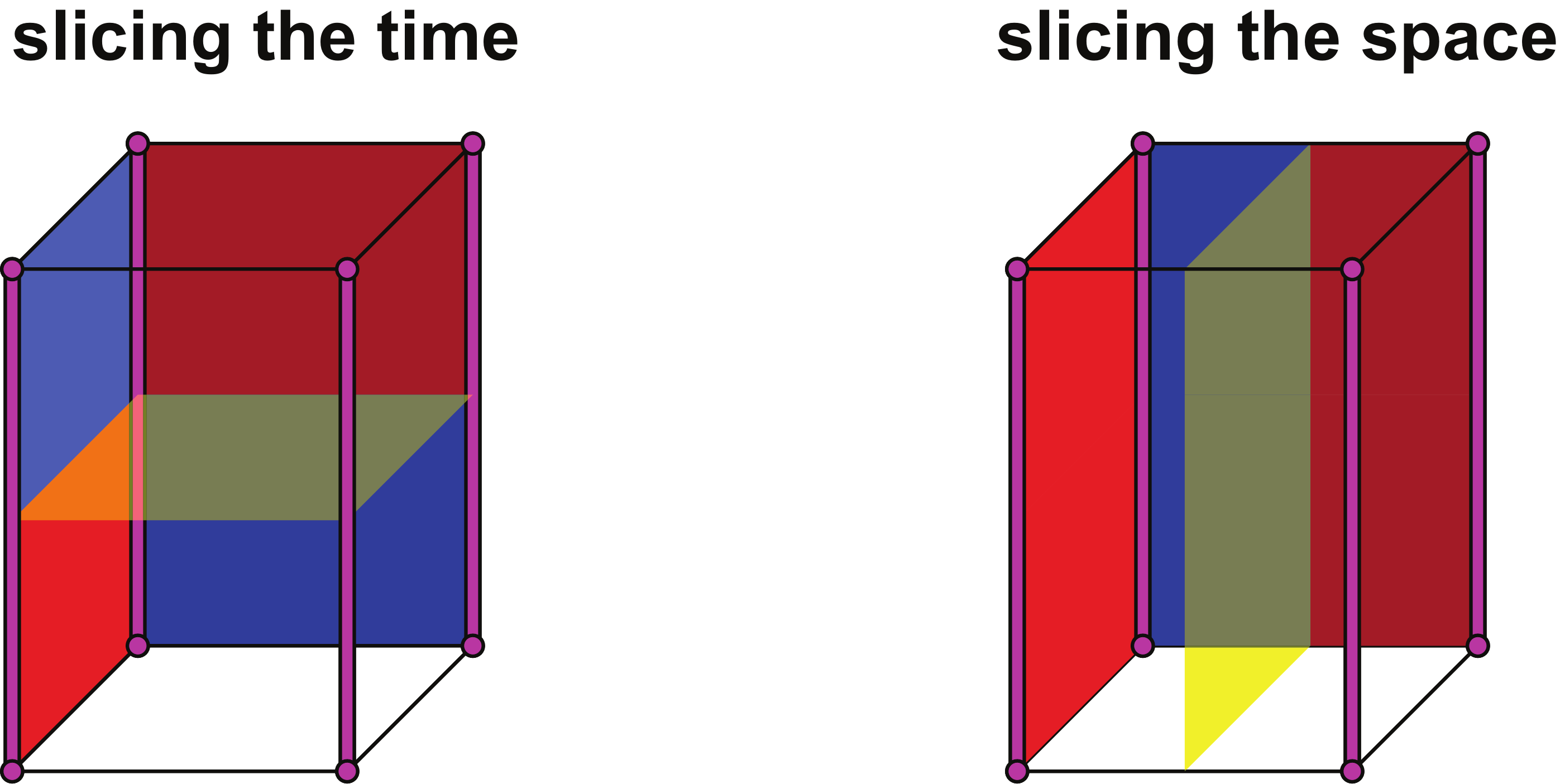}
    \caption{Defect diagrams including domain walls, depicted as a transparent yellow membrane.}
    \label{fig:domain_wall}
\end{figure}

Some code deformation protocols apply H gates on several or all data qubits, and temporarily introduce XXZZ checks~\cite{litinski2018lattice,geher2024error,yoder2017surface}. 
These operations effectively exchange X- and Z-type error chains, while Y-type error chains remain unchanged. 
In 3D defect diagrams, this exchange boundary is represented as a transparent yellow membrane, referred to as a \textit{domain wall}. 
The transversal H gate corresponds to a domain wall slicing through time, as shown in Fig.~\ref{fig:domain_wall}~(Left), allowing time-like errors to pass through it. 
Alternatively, by introducing XXZZ checks, we can implement a domain wall slicing through space, as shown in Fig.~\ref{fig:domain_wall}~(Right), allowing space-like errors to pass through it. 

To validate that a code deformation procedure implements a target unitary Clifford operation $C$, one can track the evolution of logical Pauli operators. 
Specifically, an initial logical Pauli operator $P$ must map to $CPC^{\dagger}$ after the deformation. 
For example, in the case of logical S gates, we should check the logical X operator maps to the logical $SXS^{\dagger}=Y$ operator (i.e., X$\rightarrow$Y), while the logical Z operator remains unchanged (Z$\rightarrow$Z), up to multiplication by stabilizers.

\section{Motivation and Existing methods}\label{sec:motivation}
To accelerate the computation of fault-tolerant quantum computing with surface codes, efficient implementations of universal logical gate sets are required. 
Typically, we choose H, S, CNOT, T as a basic unitary gate set and focus on the implementation of them. 
Among these operations, T gates were historically considered the most expensive because they involve magic-state distillation. 
However, recent protocols, such as magic-state cultivation~\cite{gidney2024magic}, have significantly reduced this cost. 
Thus, it is strongly demanded to reduce the cost of other Clifford operations so that they do not become a bottleneck.

This paper focuses on the implementation of logical S gates. 
An implementation based on code deformation was originally proposed in Ref.~\cite{brown2017poking} by exchanging the locations of two adjacent twist defects via braiding. 
However, that work provided only defect diagrams and did not show circuit-level implementation, such as detector diagrams. 
Thus, efficiently implementing this operation while maintaining the fault distance remained an open problem. 
Since then, several studies have addressed this issue. 
In this section, we review two recent approaches that provide an explicit implementation of logical S gates.

In the following discussion, we assume that all physical two-qubit gates have identical durations and can be executed simultaneously. 
While this assumption may not hold in practical hardware implementations and requires careful consideration, addressing such hardware-specific constraints is beyond the scope of this work.

\subsection{Bombín's methods~\cite{bombin2023logical}}
Bombín {\it et al.}~\cite{bombin2023logical} introduced the first explicit implementation for rotated surface codes, which we refer to as Bombín's method. 
The 2D and 3D defect diagrams for this method are shown in Fig.~\ref{fig:bombin_schematic} and Fig.~\ref{fig:defect_diagram_bombin}, respectively. 
This method requires a spacetime volume of $2d \times 2d \times d$. 
The procedure in the figure is as follows:

\begin{figure}[t]
    \centering
    \includegraphics[scale=0.27]{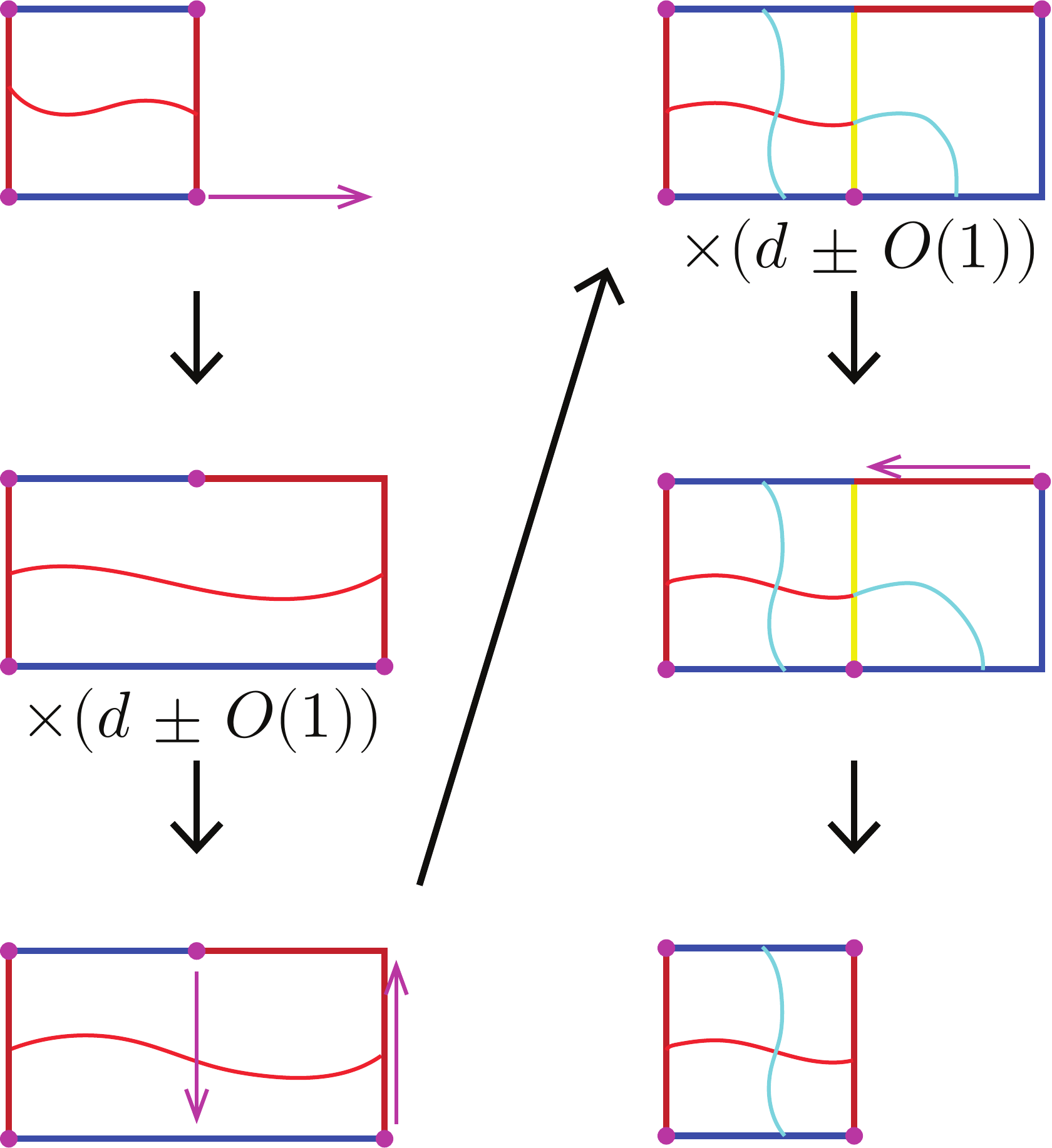}
    \caption{2D defect diagram for Bombín's method.}
    \label{fig:bombin_schematic}
\end{figure}
\begin{figure}[t]
    \centering
    \includegraphics[scale=0.1]{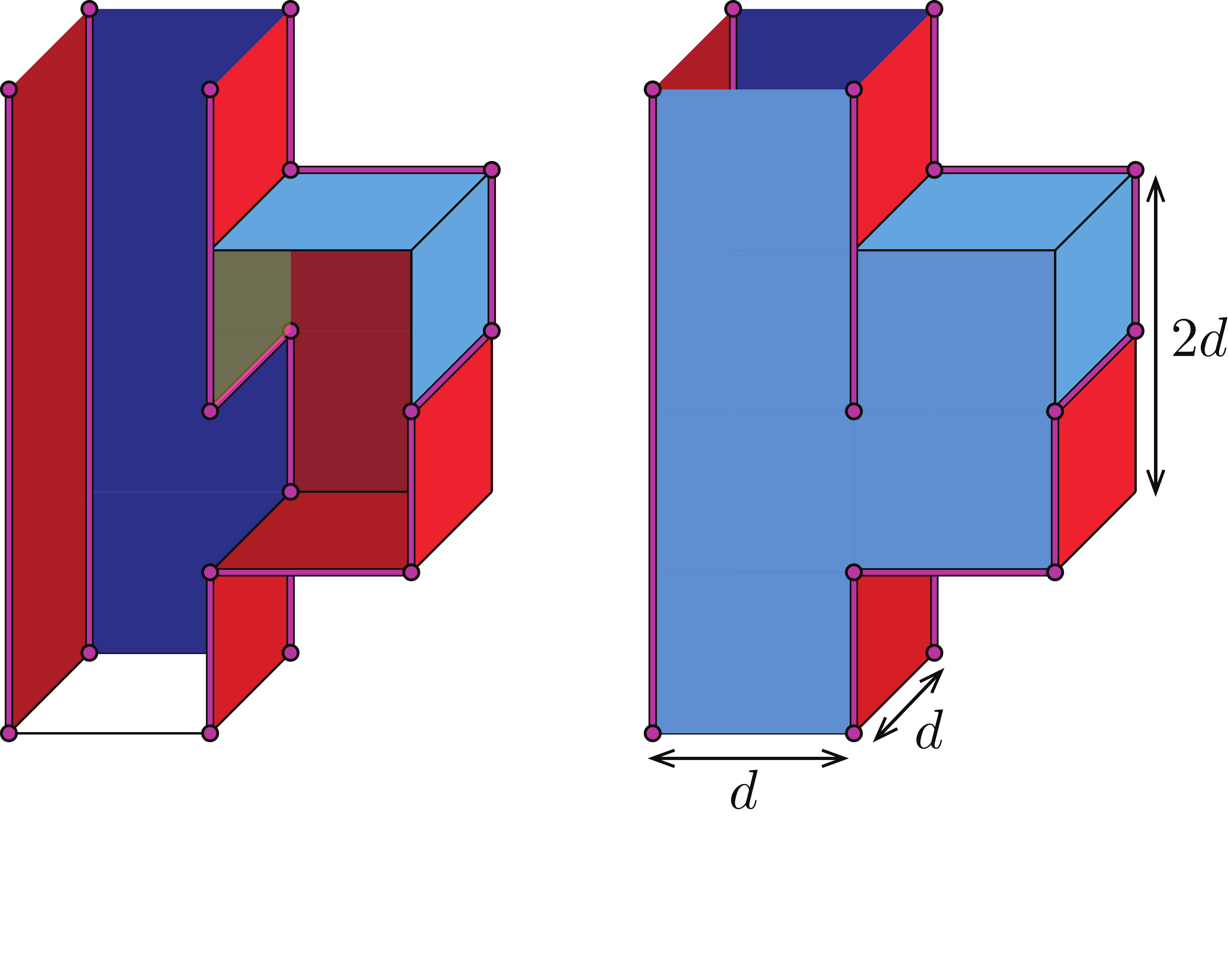}
    \vspace{-35pt}
    \caption{3D defect diagram of Bombín's method. The time goes upward.}
    \label{fig:defect_diagram_bombin}
\end{figure}

\begin{enumerate}
    \item Expand the surface code spatially from the X boundary while braiding the twist defects in the middle-bottom to the right-bottom corner.
    \item Perform $d \pm O(1)$ rounds of syndrome measurement to maintain the fault distance $d$.
    \item Braid the twist defects from the middle-top to the bottom side, and from the right-bottom to the right-top corner.     
    % \item Perform a transversal H gate on the right half to create a domain wall.
    \item Perform $d \pm O(1)$ rounds of syndrome measurement again to maintain the fault distance $d$.
    \item Shrink the surface code to its original size.
\end{enumerate}
Note that while we describe Bombín's method with a domain wall slicing through time, it can also be implemented with a domain wall slicing through space by utilizing XXZZ stabilizers in the middle column.
To the best of our knowledge, this is the most efficient implementation that maintains a separation of at least $d$ between boundaries and defect pairs at all stages of the code deformation.

\begin{figure}[t]
    \centering
    \includegraphics[scale=0.22]{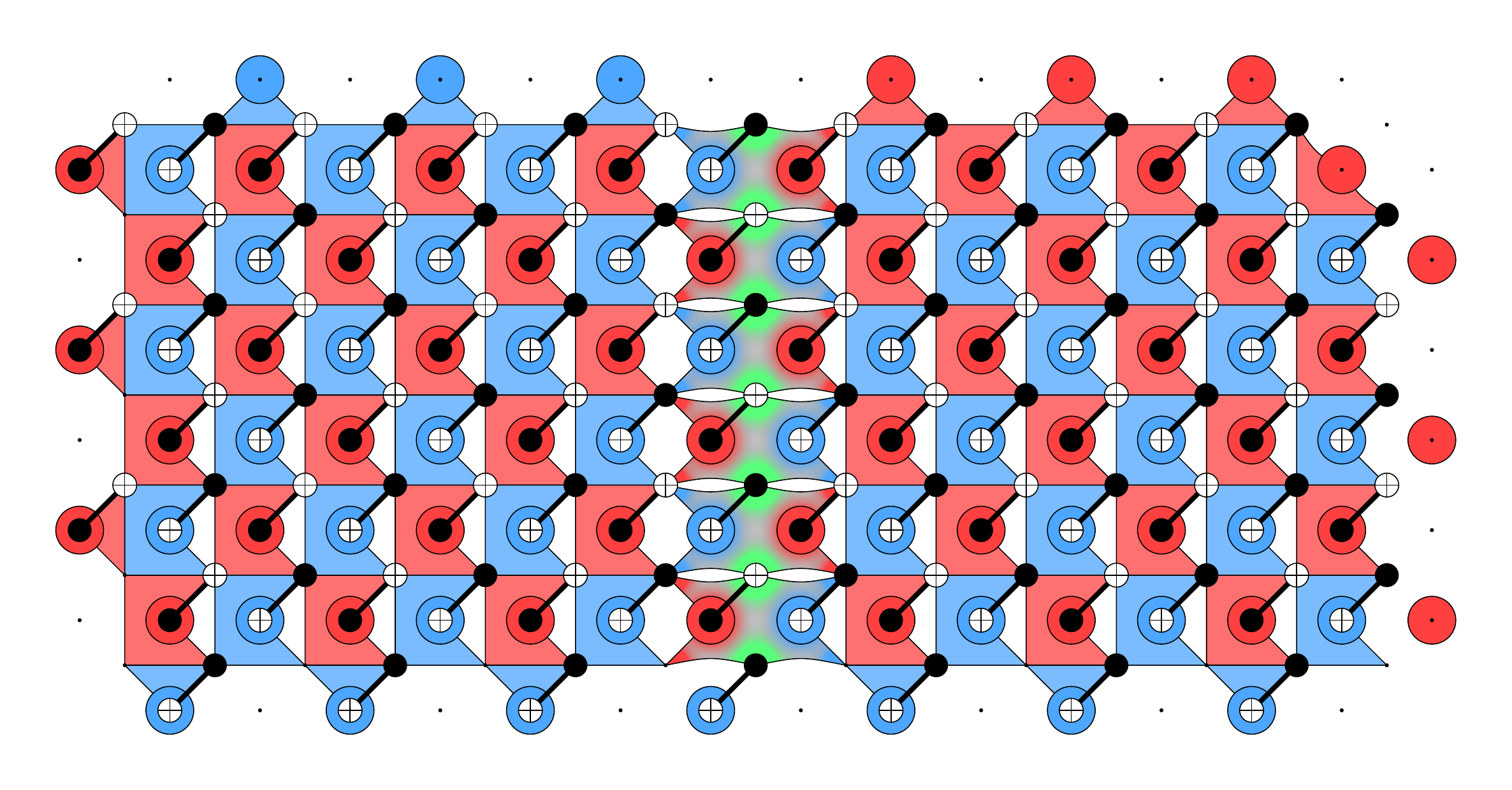}
    \caption{Stim circuit diagram for Bombín's method at the moment when the twist defect is braided across the surface code.}
    \label{fig:bombin_braiding}
\end{figure}

A major drawback in Bombín's method is that SWAP gates are required after the twist defect braiding through the bulk of the surface code. 
This step is necessary to transform the resulting rectangular stabilizers (shown in Fig.~\ref{fig:bombin_braiding}) back into standard square ones, a requirement imposed by nearest-neighbor connectivity constraints~\cite{bombin2023logical}. 
%The rectangular stabilizers are shown in Fig.~\ref{fig:bombin_braiding}. 
This transformation introduces additional two-qubit gate depth, making synchronization with other surface code patches difficult~\cite{maurya2025synchronization} and inducing additional idling errors on the data qubits. 
Furthermore, these rectangular stabilizers exhibit redundant Y-type support on data qubits, rendering both Z- and X-type hook errors unavoidable in schemes employing a four-layer two-qubit gate depth for syndrome extraction. 
Consequently, the upper bound of the fault distance is effectively reduced to $d/2$ as we demonstrate later in Sec.~\ref{sec:results}.

\subsection{Gidney's methods~\cite{gidney2024inplace}}
\begin{figure*}[t]
    \centering
    \includegraphics[scale=0.30]{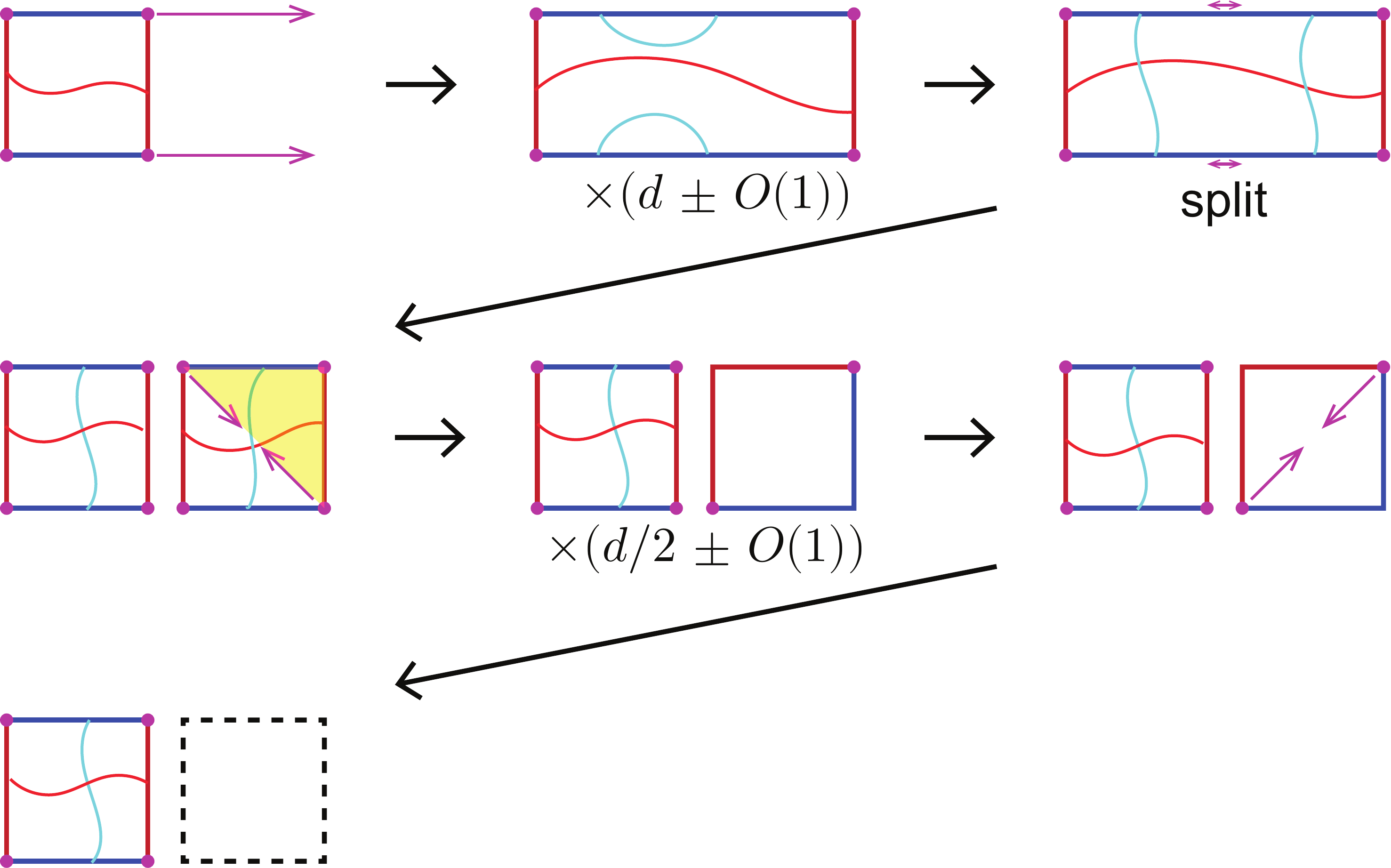}
    \caption{2D defect diagram of Gidney's method.}
    \label{fig:gidney_schematic}
\end{figure*}

\begin{figure}[t]
    \centering
    \includegraphics[scale=0.1]{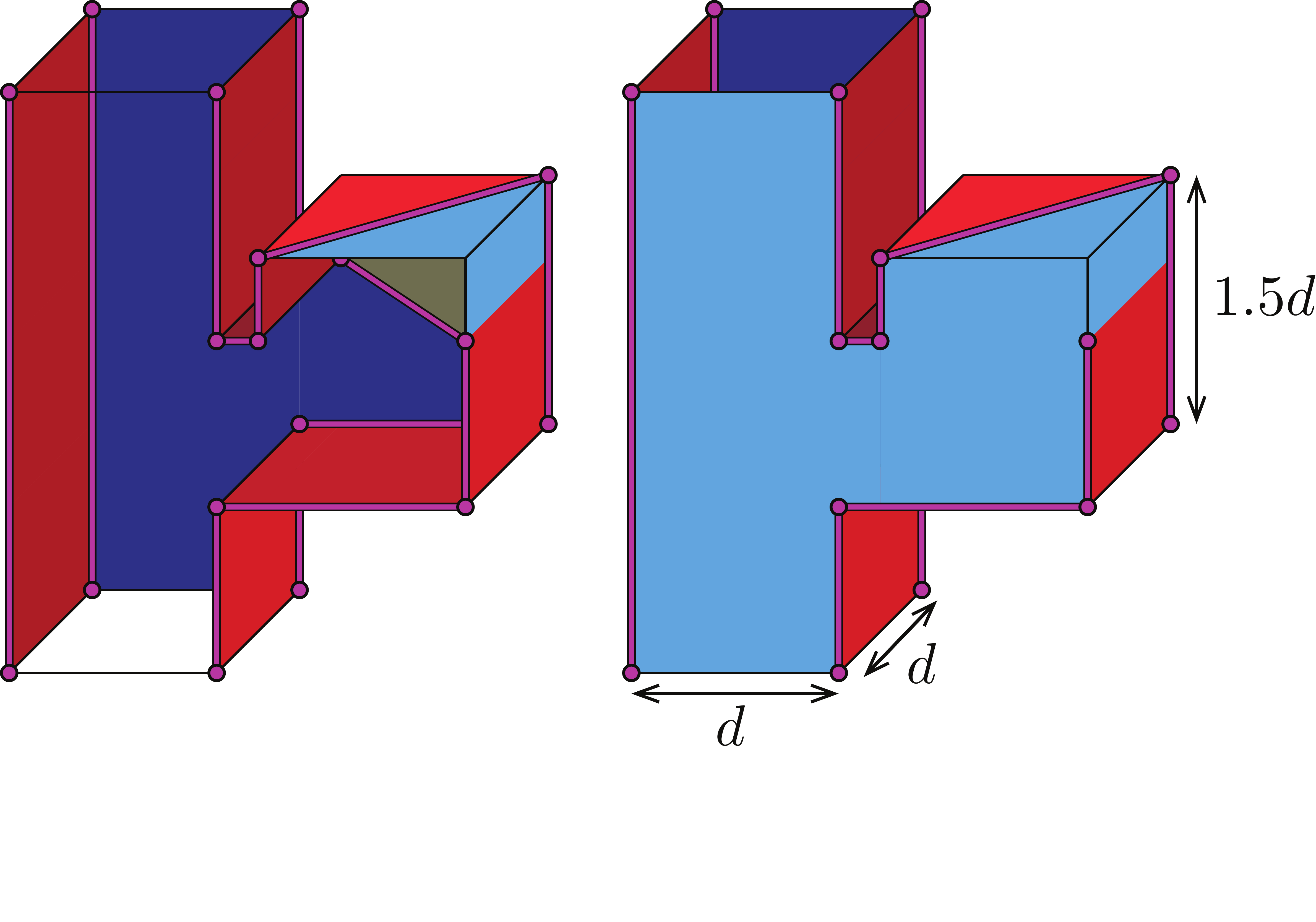}
    \vspace{-35pt}
    \caption{3D defect diagram of Gidney's method. The time goes upward.}
    \label{fig:defect_diagram_gidney}
\end{figure}

Gidney~\cite{gidney2024inplace} provided a more spacetime-efficient implementation of the logical S gate in surface codes, which we call Gidney's method. 
Its defect diagrams are shown in Fig.~\ref{fig:gidney_schematic} and Fig.~\ref{fig:defect_diagram_gidney}.
This method requires a spacetime volume of $2d \times 1.5d \times d$. 
The procedure in the figure is as follows:
\begin{enumerate}
    \item Expand the surface code spatially from the X boundary.
    \item Perform $d \pm O(1)$ rounds of syndrome measurement to maintain the fault distance $d$.
    \item Split the surface code into two patches via lattice surgery.
    \item Perform an inplace Y measurement on the right surface code patch.
\end{enumerate}

At a glance, Gidney's implementation does not appear to maintain a full fault distance $d$, as the pair of twist defects is separated by only $d/2$ geometrically. 
However, due to the asymmetry of the Y-error mechanism between time-like and space-like errors, it was shown that Y-errors cannot grow along the time direction with a chain of only $d/2$ errors; that is, they cannot directly connect one defect to the other. 
Thus, the effective fault distance of Gidney's implementation remains $d$.

The right panel of Figure 9 in Ref.~\cite{gidney2024inplace} presents a construction of the logical S gate with a minimal spacetime volume.
However, this applies only to moving logical qubits and is unsuitable for stationary logical qubits due to the aforementioned Y-error asymmetry. 
Therefore, to the best of our knowledge, the most volume-efficient implementation of the logical S gate for a stationary logical qubit is depicted in the left panel of Figure 9 in Ref.~\cite{gidney2024inplace}, and it remains an open question whether the volume can be further reduced.

Unlike Bombín's method, no rectangular stabilizers arise in Gidney's method. 
However, another concern in Gidney's method~\cite{gidney2024inplace} is that the logical Y measurement on the square surface code inherently requires a two-qubit gate depth of five (see Figure 7 in Ref.~\cite{gidney2024inplace}), which introduces additional idling errors on the data qubits. 
Nevertheless, synchronization is not problematic in this case because the surface code patch is measured afterward. 

\subsection{Motivation}
Summarizing the existing proposals, the achievable spacetime volume for logical S gates is $2d \times 2d \times d$ when keeping boundaries and defects apart by a distance of $d$. 
With a careful analysis of Y-type error asymmetry, the volume can be reduced to $2d \times 1.5d \times d$, which motivates the search for even more efficient implementations. 
If we are allowed to impose additional constraints, such as moving logical qubits, the volume can be reduced to $1.5d \times d \times d$; however, this is not applicable to stationary logical qubits. 
It remains an open question whether we can reduce the spacetime volume below $2d \times 1.5d \times d$ while maintaining a comparable logical error rate. As explained in the next section, we will show a $2d \times d \times d$ implementation of logical S gates for stationary logical qubits, keeping a reduction of the fault distance constant against the code distance.

Another vital existing problem would be that there are no publicly available circuit-level implementations of logical S gates. 
This makes it difficult to quantitatively compare the proposals with the existing implementation using standard libraries, such as Stim~\cite{gidney2021stim}. 
For example, as we will show later, Bombín's implementation actually does not maintain fault distance of $d$, although it exhibits logical error rate suppression similar to that of a fault-distance-$d$ implementation. This is difficult to check only from theoretical proposals. 
Thus, providing explicit implementation on a commonly utilized platform is vital for future studies. 
We have implemented not only our proposed schemes but also the two existing schemes with Stim, and made them publicly available in the GitHub repository~\cite{stimcircuit_github}.

\begin{figure*}[t]
    \centering
    \includegraphics[scale=0.32]{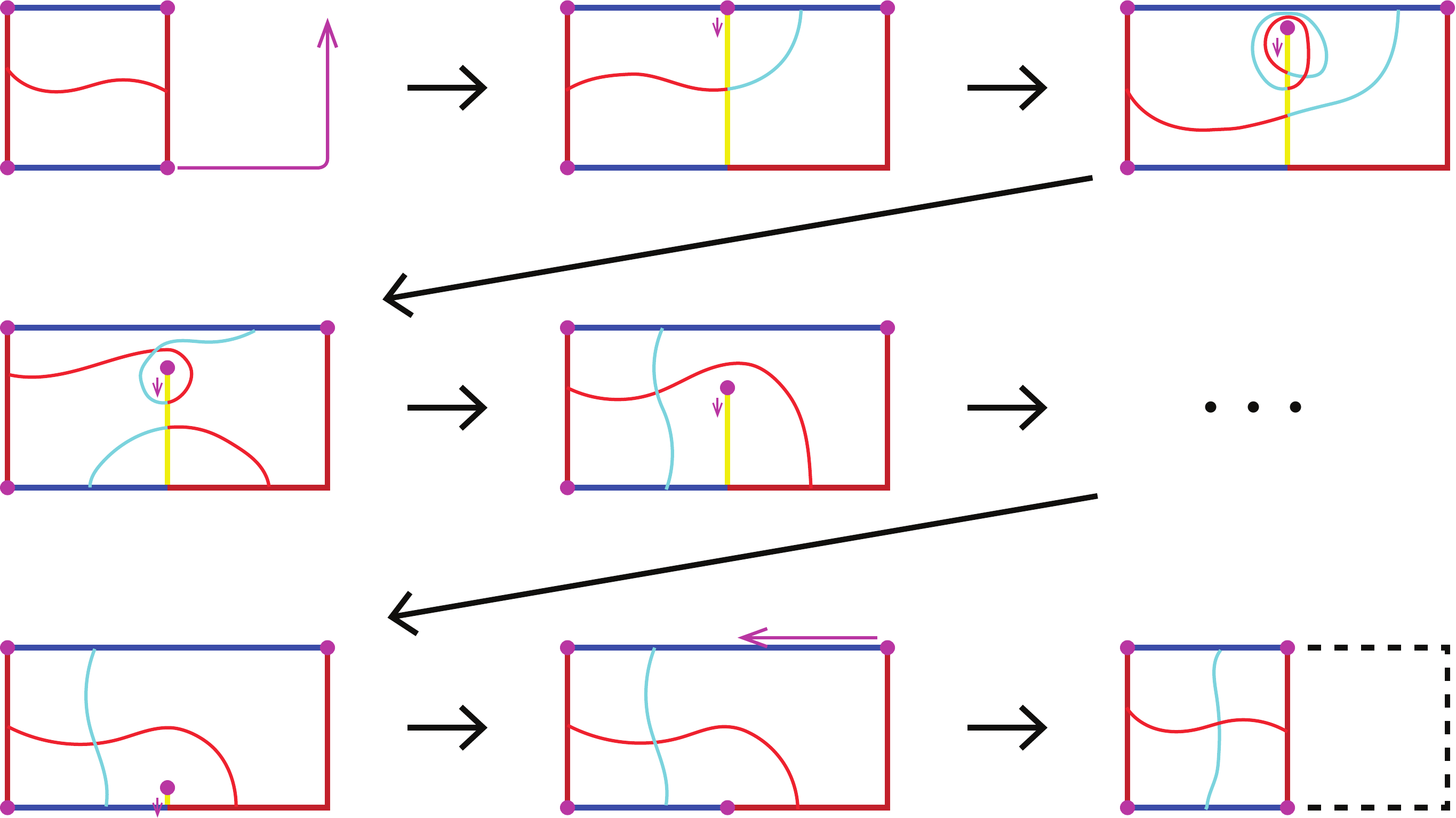}
    \caption{2D defect diagram of the proposed method.}
    \label{fig:proposed_schematic}
\end{figure*}

\begin{figure}[t]
    \centering
    \includegraphics[scale=0.11]{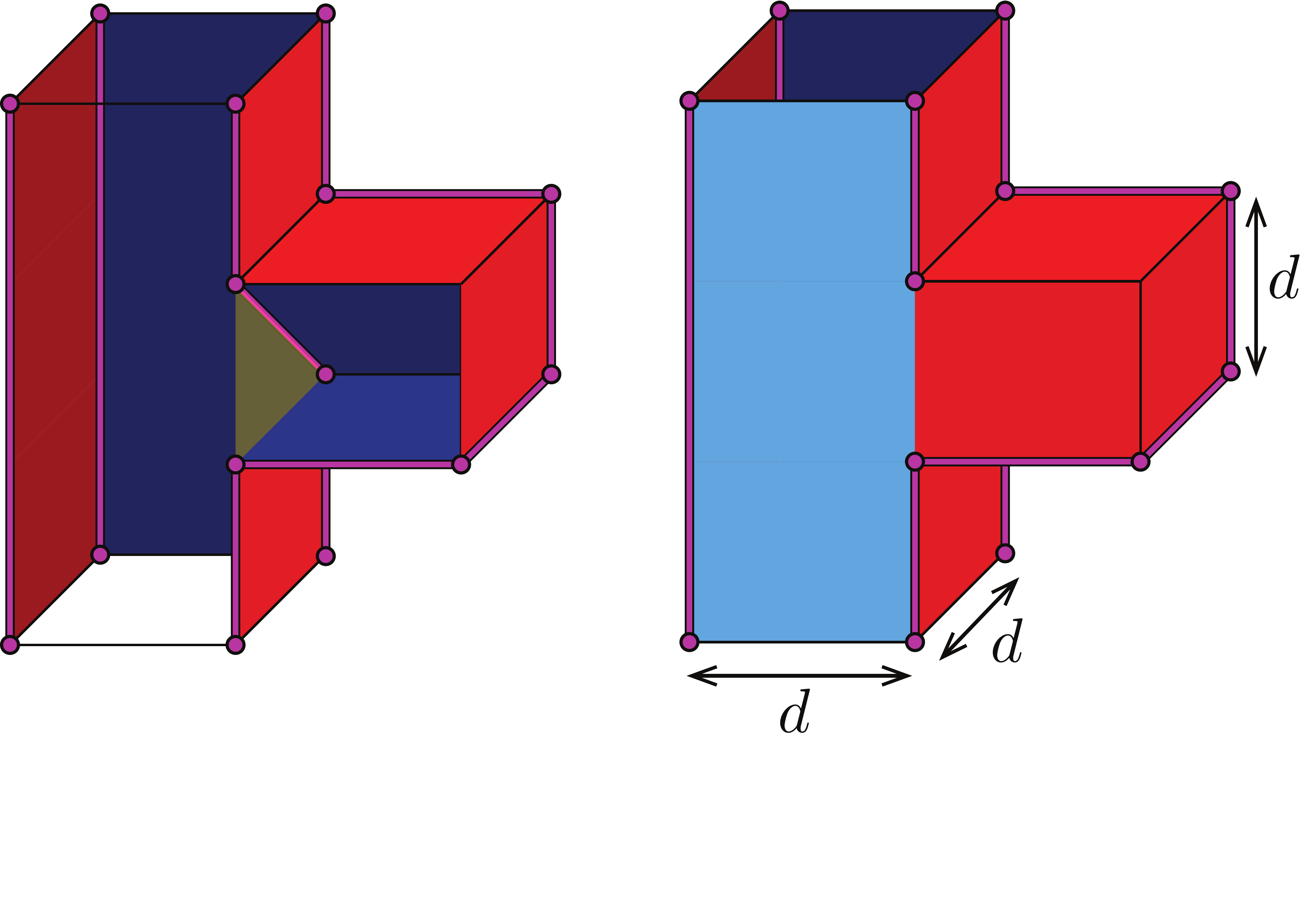}
    \vspace{-35pt}\caption{3D defect diagram of the proposed method. One of the twist defects braided through the bulk of the surface code in both space and time directions. The time goes upward.}
    \label{fig:defect_diagram_proposed}
\end{figure}

\begin{figure}[t]
    \centering
    \includegraphics[scale=0.12]{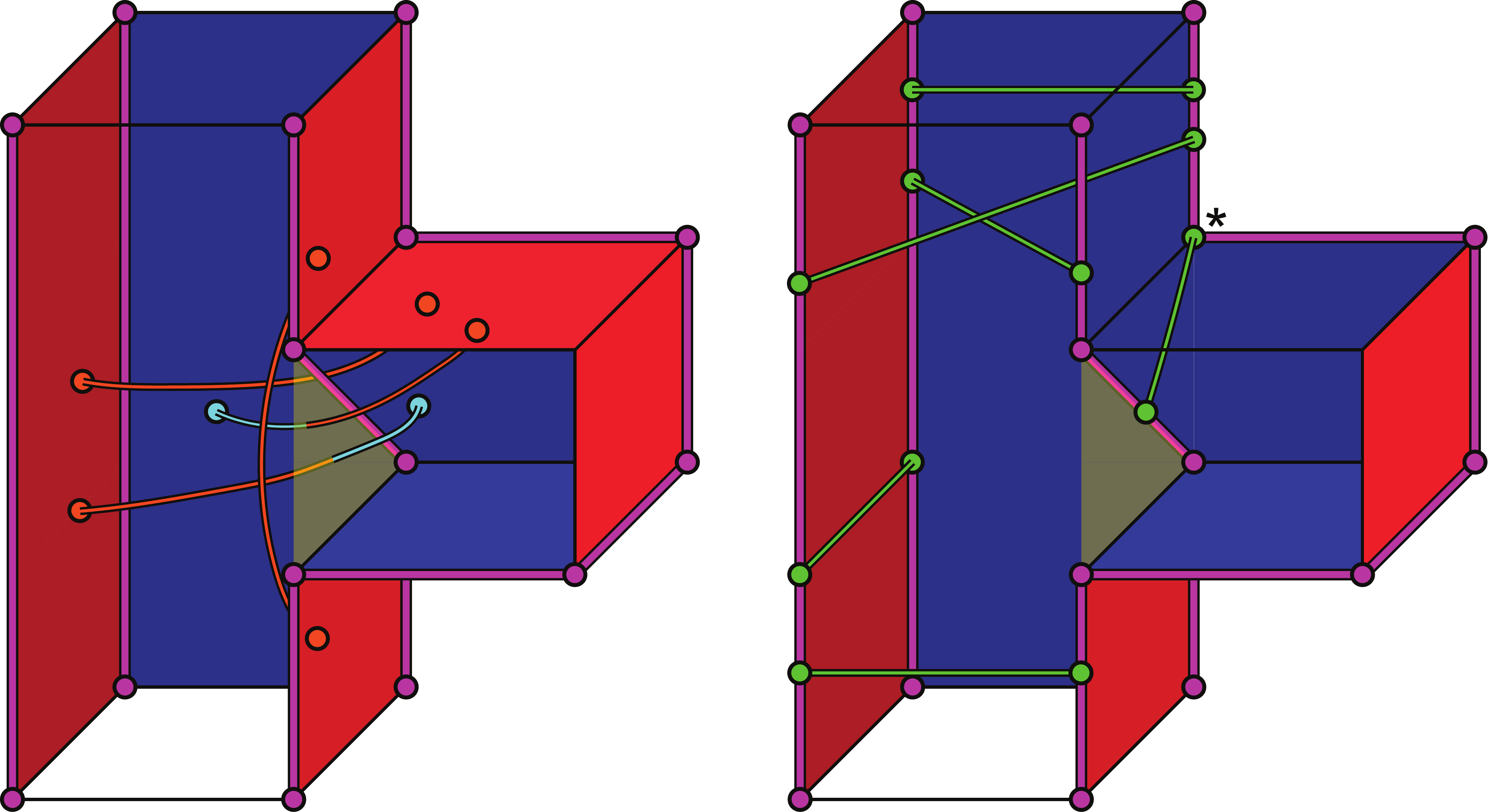}
    \caption{The error chains in the proposed method. The blue (red) strings represent Z (X) error chains in the left panel. The green strings represent Y error chains in the right panel.}
    \label{fig:error_chains_proposed}
\end{figure}

\begin{figure*}[t]
    \centering
    \includegraphics[scale=0.18]{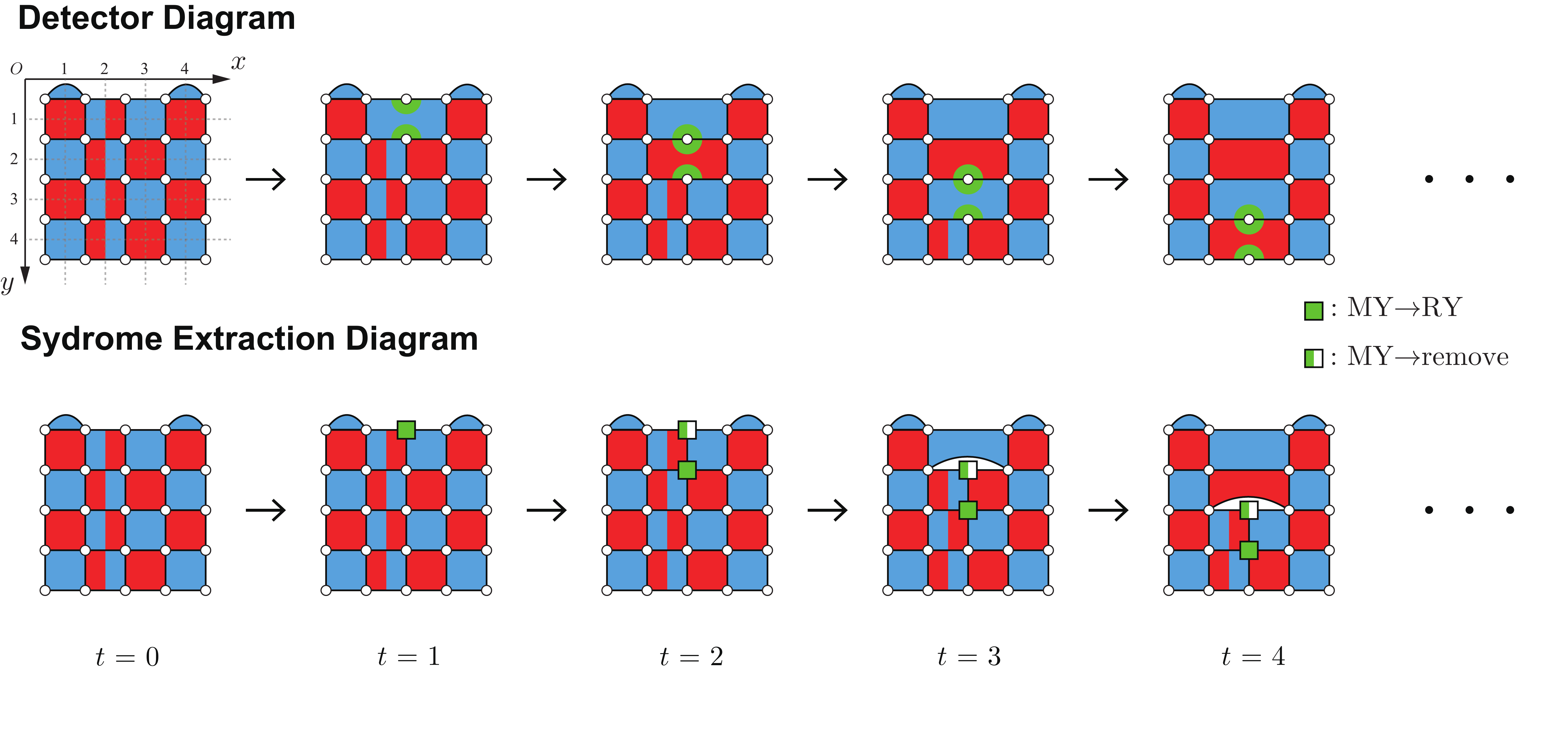}
    \vspace{-40pt}
    \caption{Detector and syndrome extraction diagrams for the proposed method using some constant-length non-local gates. The detector diagram specifies the stabilizer generators, while the syndrome extraction diagram describes the measurement patterns used for syndrome extraction. In the syndrome extraction diagram, the measurements are allowed to mutually anticommute.}
    \label{fig:diagram_proposed_non_local}
\end{figure*}

\section{Proposed Method}\label{sec:proposed}

\begin{figure}[t]
    \centering
    \includegraphics[scale=0.12]{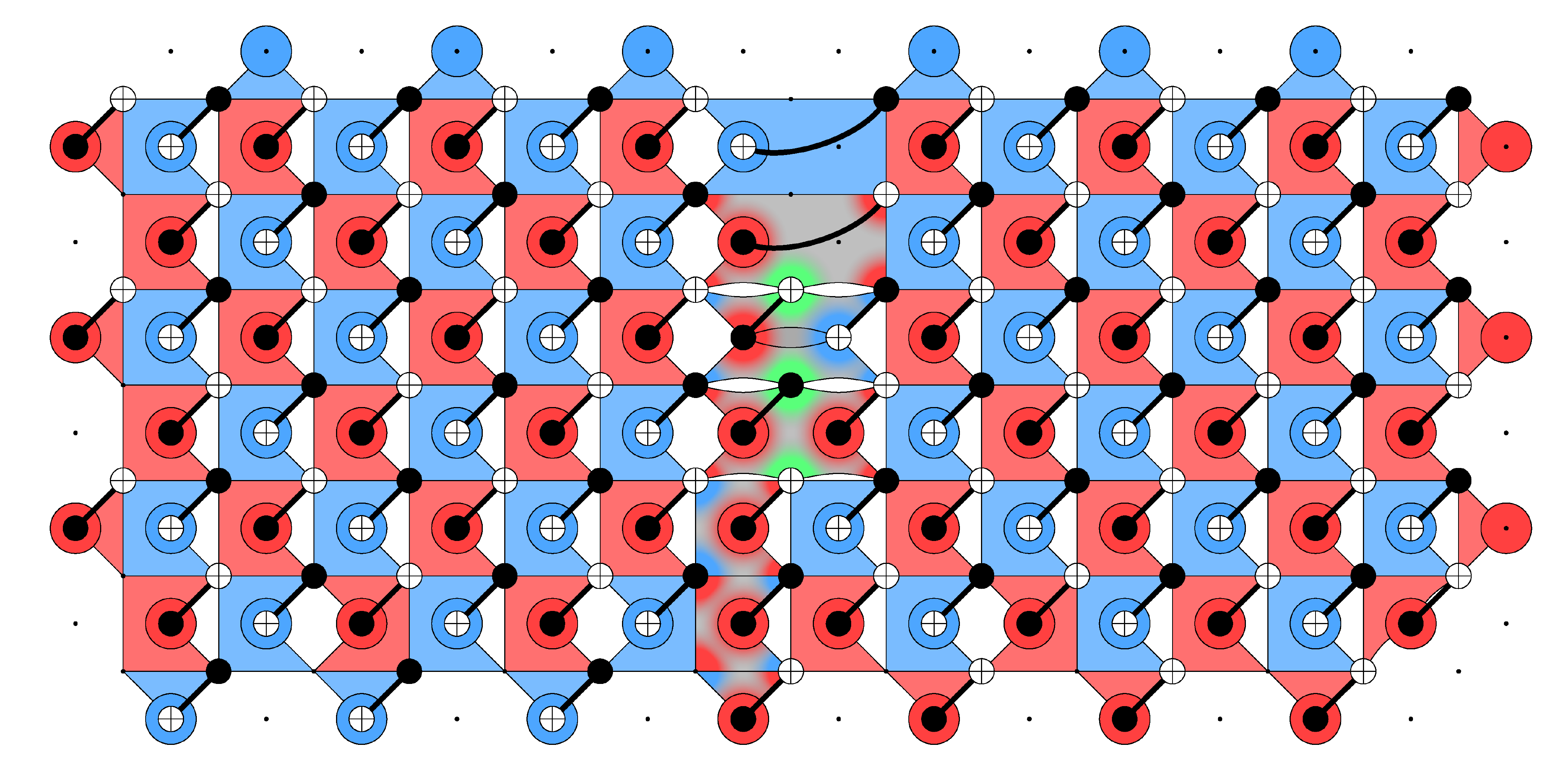}
    \caption{Stim circuit for the proposed method at the moment when the twist defect is braided across the surface code.}
    \label{fig:circuit_proposed_non_local}
\end{figure}

\begin{figure*}[t]
    \centering
    \includegraphics[scale=0.18]{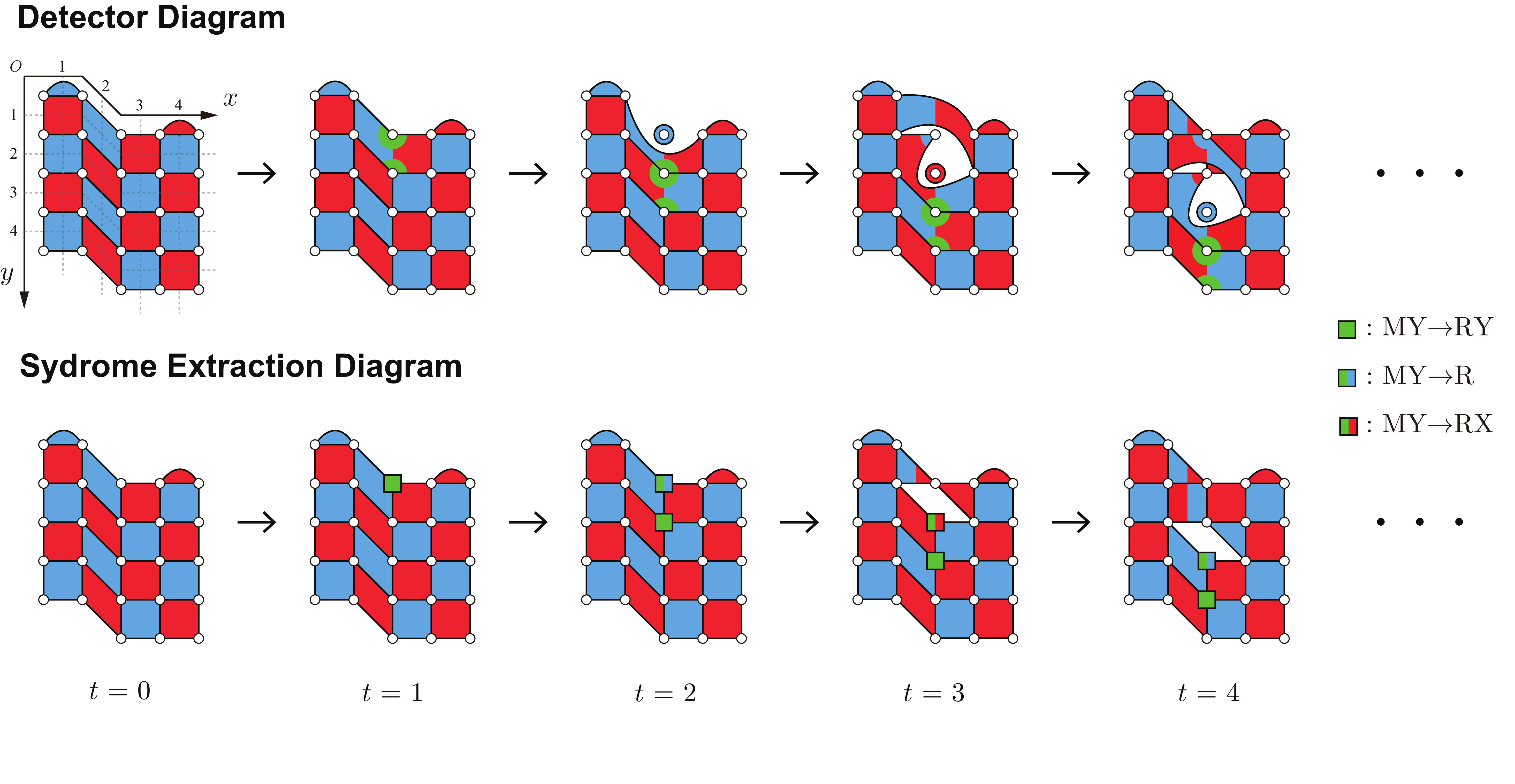}
    \vspace{-40pt}
    \caption{Detector and syndrome extraction diagrams for the proposed method using only local gates.}
    \label{fig:diagram_proposed_local}
\end{figure*}

In this section, we provide a $2d \times d \times d$ implementation of our method. In the construction phase, we utilized Crumble~\cite{gidney_crumble} and Stim~\cite{gidney2021stim}. Here, instead of presenting the Crumble example, we outline the design for the proposed method.

\subsection{Defect Diagram}
The 2D defect diagram of the proposed method is shown in Fig.~\ref{fig:proposed_schematic}. The procedure is as follows:
\begin{enumerate}
    \item Expand the surface code spatially from the X boundary while braiding the middle-bottom twist defect to the right-top corner with creating the domain wall.
    \item Perform sequential Y measurements along the middle column to gradually move the middle-top twist defect to the bottom side.
    \item Shrink the surface code to its original while braiding the twist defect at the right-top corner to the middle-top.
\end{enumerate}
Note that while Gidney's method utilizes the triangular domain wall slicing through time, the proposed method uses the triangular domain wall slicing through space using the XXZZ stabilizers in the middle column.
The 3D defect diagram of the proposed method is shown in Fig.~\ref{fig:defect_diagram_proposed}. Consequently, the method requires a spacetime volume of $2d \times d \times d$. Additionally, the logical errors that will occur in the proposed method are shown in Fig.~\ref{fig:error_chains_proposed}. The blue (red) strings represent Z (X) error chains in the left panel. The green strings represent Y error chains in the right panel. The strings can be deformed topologically within the diagram with a constraint that the endpoints of the blue and red strings remain on the walls of the same color, and those of the green strings remain on the purple lines. Note that the domain wall flips the type between Z and X. As illustrated in the left panel, all error chains must have a length of at least $d$. In contrast, in the right panel, the shortest error chain, denoted by ``*'', appears to have a Euclidean distance of $\sqrt{2}d/2$. One might be concerned that this would reduce the fault distance to $\sqrt{2}d/2$. However, this is not the case. The growth of the Y-type error chains is asymmetric between space and time~\cite{gidney2024inplace}; specifically, more than $d - 1$ Y-type errors are required to form such a logical error. The others apparently have a distance of at least $d$.

\subsection{Implementation with some constant-length non-local gates}

To implement the proposed method using some constant-length non-local gates, we provide a bit more detailed explanation in Fig.~\ref{fig:diagram_proposed_non_local}. The detector diagram represents the stabilizer generators of the surface code supported only on the data qubits; therefore, measurement qubits are omitted. In contrast, the syndrome extraction diagram represents the actual syndrome extraction pattern and may include measurements that anti-commute with each other. Here, we assume that syndrome extraction is performed using native multi-Pauli product measurements for simplicity. To clearly identify each tile in the diagram, we introduce a coordinate system, which allows us to specify precisely which stabilizers or syndromes are being referred to in the sentences. The coordinates are denoted by $(x, y)$ in the following. In addition, we denote a point at time $t$ and coordinate $(x, y)$ as $(x, y, t)$. 

At $t = 0$, the stabilizers are in their standard configuration, and syndrome extraction is performed normally. At $t = 1$, in addition to the standard syndrome extraction as in the previous step, a single Y measurement and a single Y reset are performed at $(2.5, 0.5)$. However, the tile-like stabilizers at $(2,1)$ and $(3,1)$ are no longer individually preserved; instead, only their product, i.e., a hexagonal stabilizer, is maintained. In this step, we can construct a detector corresponding to such a stabilizer by taking the XOR of the records at $(2,1,1)$, $(3,1,1)$, $(2,1,0)$, and $(3,1,0)$. At $t = 2$, in addition to the standard syndrome extraction as in the two previous steps, two single Y measurements are performed at $(2.5, 0.5)$ and $(2.5, 1.5)$, followed by a removal operation and a Y reset, respectively. Similar to the case at $t = 1$, instead of preserving the individual stabilizers at $(2,2)$ and $(2,3)$, their product is preserved as a hexagonal stabilizer. The hexagonal stabilizer formed at $t = 1$ is inherently preserved in this step; however, we remove its support at $(2.5, 0.5)$ by taking the product with the Y measurement at $(2.5, 0.5)$. In this step, we can construct detectors for the pentagonal and hexagonal stabilizers by taking the XOR of the records at $(2,2,2)$, $(3,2,2)$, $(2,2,1)$, and $(3,2,1)$, and by taking the XOR of the records at $(2,1,2)$, $(3,1,2)$, $(2,1,1)$, $(3,1,1)$, and $(2.5,0.5,1)$, respectively. At $t = 3$, we perform a modified syndrome extraction, different from the previous steps, as depicted in the bottom panel of the figure, followed by a Y-measurement-then-Y-reset and a Y-measurement-then-removal operation, both shifted downward compared to the case at $t = 2$. New hexagonal and pentagonal stabilizers are constructed similarly to the previous cases, and the corresponding detectors are also constructed in the same manner. Here, we provide an explanation only for the middle-top stabilizer, which is obtained by taking the XOR of the record corresponding to the middle-top stabilizer at $t = 3$ with the records at $(2,1,2)$, $(3,1,2)$, $(2.5,0.5,2)$, and $(2.5,1.5,2)$. At $t = 4$, the syndrome extraction is performed in a similar way to the previous steps. New hexagonal and pentagonal stabilizers and their corresponding detectors are also constructed in the same manner as in the previous cases. The rectangular stabilizer is constructed by removing the Y-type support of the pentagonal stabilizer in the previous step by taking the product with the Y measurement at $(2.5,1.5)$. Note that such a rectangular stabilizer coincides with a weight-4 stabilizer of the surface code; therefore, its detector can be constructed straightforwardly, as in the conventional surface code. The same procedure is repeated in the subsequent steps.

The proposed method can be implemented straightforwardly by allowing constant-length non-local gates, as shown in Fig.~\ref{fig:circuit_proposed_non_local}, to extract syndromes for the rectangular stabilizers after twist-defect braiding. As shown below, the fault distance is upper-bounded by $d-1$ when such non-local gates are used. This protocol relies on constant-length non-local gates and is therefore well suited to hardware platforms that support non-nearest-neighbor interactions, such as neutral atoms. However, in superconducting devices, implementing non-nearest-neighbor interactions remains challenging. Therefore, in the following paragraph, we present a more involved implementation using only local gates, including CXSWAP gates.

\subsection{Implementation only with local gates}

The stabilizer and syndrome extraction diagrams are shown in Fig.~\ref{fig:diagram_proposed_local}. For clarity, we introduce a coordinate system whose $x$-axis is distorted to follow the geometry of the surface code. The strategy is largely based on the non-local case, but it is modified to satisfy local constraints. In the previous implementation, rectangular stabilizers remain after twist-defect braiding, requiring rectangular syndrome extraction circuits. To address this issue, we introduce the parallelogram tiles depicted in the figure. 
% By doing so, even if the rectangular stabilizers are transformed back into square ones, the tiles after twist-defect braiding align properly. 
This ensures that the stabilizer tiles remain correctly aligned after twist-defect braiding, even when the rectangular stabilizers are decomposed into square ones.
A detailed explanation follows. At $t = 0$, the stabilizers are in their standard configuration for the surface code, and syndrome extraction is performed normally. At $t = 1$, in addition to the standard syndrome extraction, a single Y measurement is performed at $(2.5,0.5)$ followed by a single Y reset. As a result, the tile-like stabilizers at $(2,1)$ and $(3,1)$ are no longer individually preserved; instead, their product is preserved as a hexagonal stabilizer. The corresponding detector is constructed by taking the XOR of the records at $(2,1,1)$, $(3,1,1)$, $(2,1,0)$, and $(3,1,0)$. At $t = 2$, syndrome extraction is performed normally, followed by Y-measurement-then-reset and Y-measurement-then-Y-reset operations. A new hexagonal stabilizer and its detector are constructed in a manner similar to the previous step. The middle-top hexagonal stabilizer in the previous step is destroyed, and a new pentagonal stabilizer is formed in this step. The corresponding detector is constructed by taking the XOR of the records at $(2,1,2)$, $(3,1,2)$, $(2,1,1)$, $(3,1,1)$, and $(2.5,0.5,1)$. At $t = 3$, the syndrome extraction pattern differs from the previous cases, as shown in the figure. The pentagonal stabilizer in the previous step is no longer preserved; instead, a quadrangular stabilizer is constructed in this step by taking the product with the Y measurement at $(2.5,1.5)$. In addition, two new hexagonal stabilizers are formed: the upper one is obtained by taking the product between the previous hexagonal stabilizer and the previous reset and the Y measurement at $(2.5,1.5)$, while the lower one is constructed similarly to the previous cases. The corresponding detectors are constructed by taking the XOR of the records $(2,1,3)$, $(3,1,3)$, $(2,1,2)$, $(3,1,2)$, $(2.5,0.5,2)$, and $(2.5,1.5,2)$ for the quadrangular stabilizer, and $(2,2,3)$, $(3,2,3)$, $(2,2,2)$, $(3,2,2)$, and $(2.5,1.5,2)$ for the upper hexagonal stabilizer; the lower one is analogous. At $t = 4$, syndrome extraction is performed as shown in the figure. The previous upper hexagonal stabilizer is no longer preserved; instead, a pentagonal stabilizer is constructed by taking the product with the Y measurement at $(2.5,2.5)$. Finally, in this step, we obtain new square and triangular stabilizers at the middle-top, so that both the left and right sides of the surface code align properly. The detectors for stabilizers involving more than four data qubits are constructed in the same manner as in the previous steps. The detectors for the triangular and square stabilizers are constructed by taking the XOR of the records at $(2,1,4)$, $(3,1,4)$, $(2,1,3)$, and $(3,1,3)$. In the subsequent steps, the stabilizers in the region through which the twist defect passes are transformed into square stabilizers, and their detectors are constructed straightforwardly, similarly to those of the conventional surface code.

\begin{figure*}[t]
    \centering
    \includegraphics[scale=0.13]{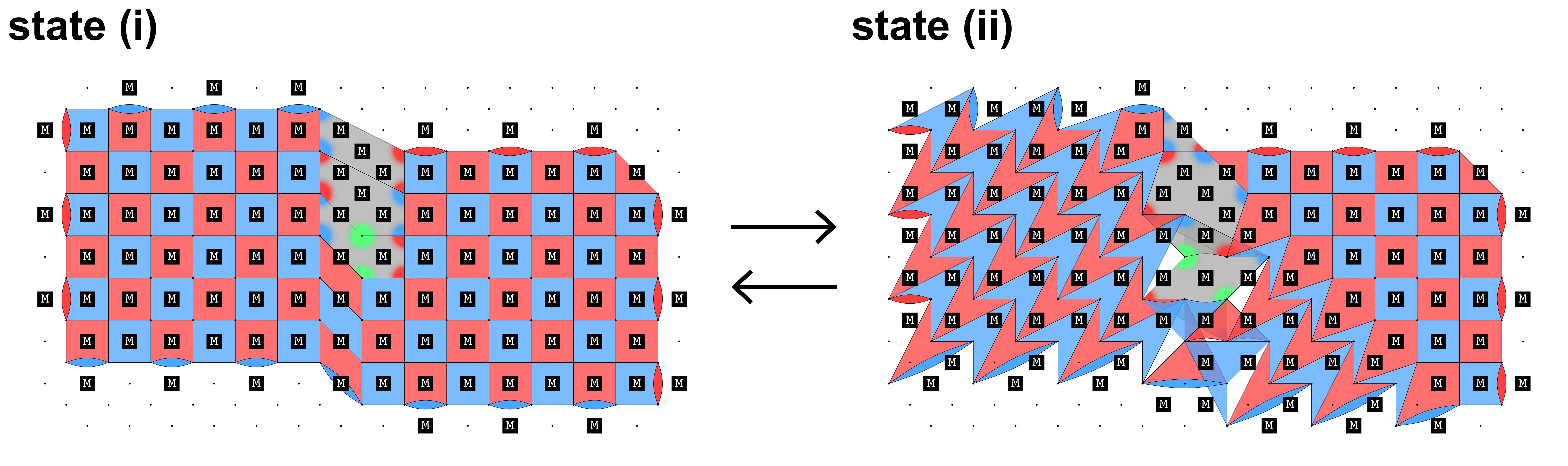}
    \caption{Partial detector diagram for implementing the proposed method using only local gates (i) before using CXSWAP gates and (ii) after using CXSWAP gates.}
    \label{fig:circuit_proposed_local}
\end{figure*}

Throughout the procedure, syndrome extraction can be performed using only local gates. Here, we present an example circuit that realizes this. The corresponding detector diagrams are shown in Fig.~\ref{fig:circuit_proposed_local}. In the circuit-level implementation, we additionally consider the ordering of two-qubit gates. As a result, the procedure differs slightly from the multi-Pauli measurement case; however, the high-level strategy remains the same. The detector configuration begins in state (i) and transforms into state (ii) by replacing specific CNOT gates with CXSWAP gates during syndrome measurement, following an approach similar to Ref.~\cite{mcewen2023relaxing}. This modification enables the implementation of parallelogram-shaped stabilizers in the middle column of (i). Once state (ii) is reached, the subsequent syndrome measurement is performed with the two-qubit gate order reversed, thereby returning the detector configuration to state (i). Consequently, the system oscillates between states (i) and (ii) as the twist defect braiding goes downward. As discussed below, the fault distance in this case is upper bounded by $d-3$.

\section{Numerical Results}\label{sec:results}
In this section, we provide the numerical results for our proposals and compare them with Bombín's and Gidney's methods. We adopted a circuit-level noise model where uniform depolarizing errors are applied to every gate and every reset, measurement, and idling operation.
\subsection{Fault Distances}

\begin{table}[t]
    \caption{Fault distance for each method}
    \label{tab:fault_distance}
    \begin{tabular}{|c|c|}
        \hline
        Method & Fault distance \\
        \hline
        \hline
        \makecell{\bf Bombín's \cite{bombin2023logical}} & $d/2$ \\
        \hline
        \makecell{\bf Gidney's \cite{gidney2024inplace}} & $d$ \\
        \hline
        \makecell{\bf Proposed (NL)\\(Fig.~\ref{fig:circuit_proposed_non_local})} & $d-1$ \\
        \hline
        \makecell{\bf Proposed (L) \\ Fig.~\ref{fig:circuit_proposed_local}}& $d-3$ \\
        \hline
    \end{tabular}
\end{table}
First, we calculated the upper-bounds of the fault distances for all methods using the Stim function \texttt{shortest\_graphlike\_error} and \texttt{search\_for\_undetectable\_logical\_errors}. The settings for the latter function are as follows:
\begin{lstlisting}[basicstyle=\ttfamily\scriptsize]
circuit.search_for_undetectable_logical_errors(
    dont_explore_detection_event_sets_with_size_above=d,
    dont_explore_edges_with_degree_above=5,
    dont_explore_edges_increasing_symptom_degree=False,
    canonicalize_circuit_errors=False,
)
\end{lstlisting}
We checked the fault distance with the former function to code distances up to 23, and the latter function to code distances up to seven.

\begin{figure*}[h]
    \centering
    \includegraphics[scale=0.12]{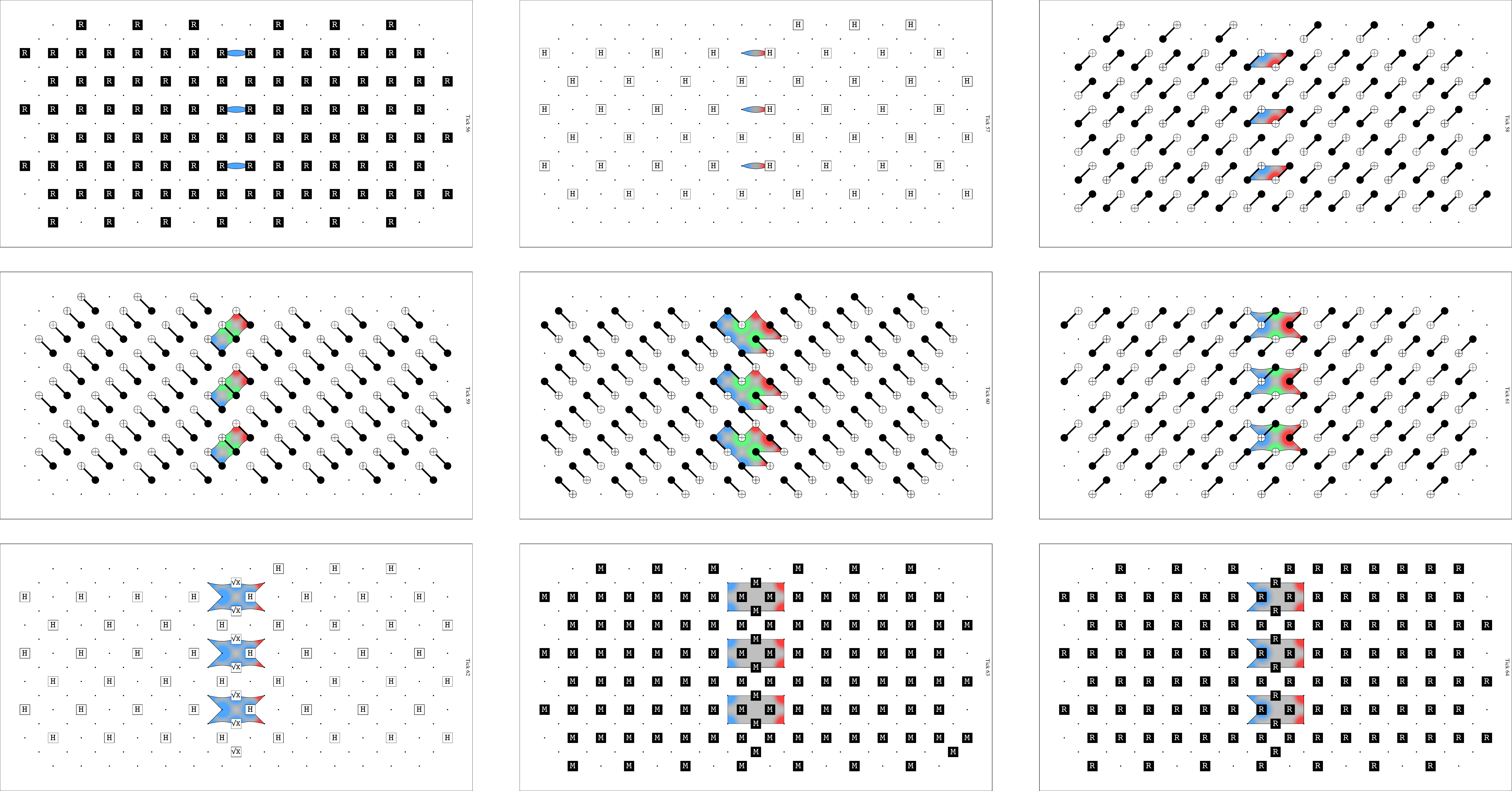}
    \caption{ The set of detectors triggered by each error in the shortest error chain that limits the fault distance to $d/2$ for Bombín's method.}
    \label{fig:circuit_detslice_bombin_hook_error}
\end{figure*}

\begin{figure*}[h]
    \centering
    \includegraphics[scale=0.1]{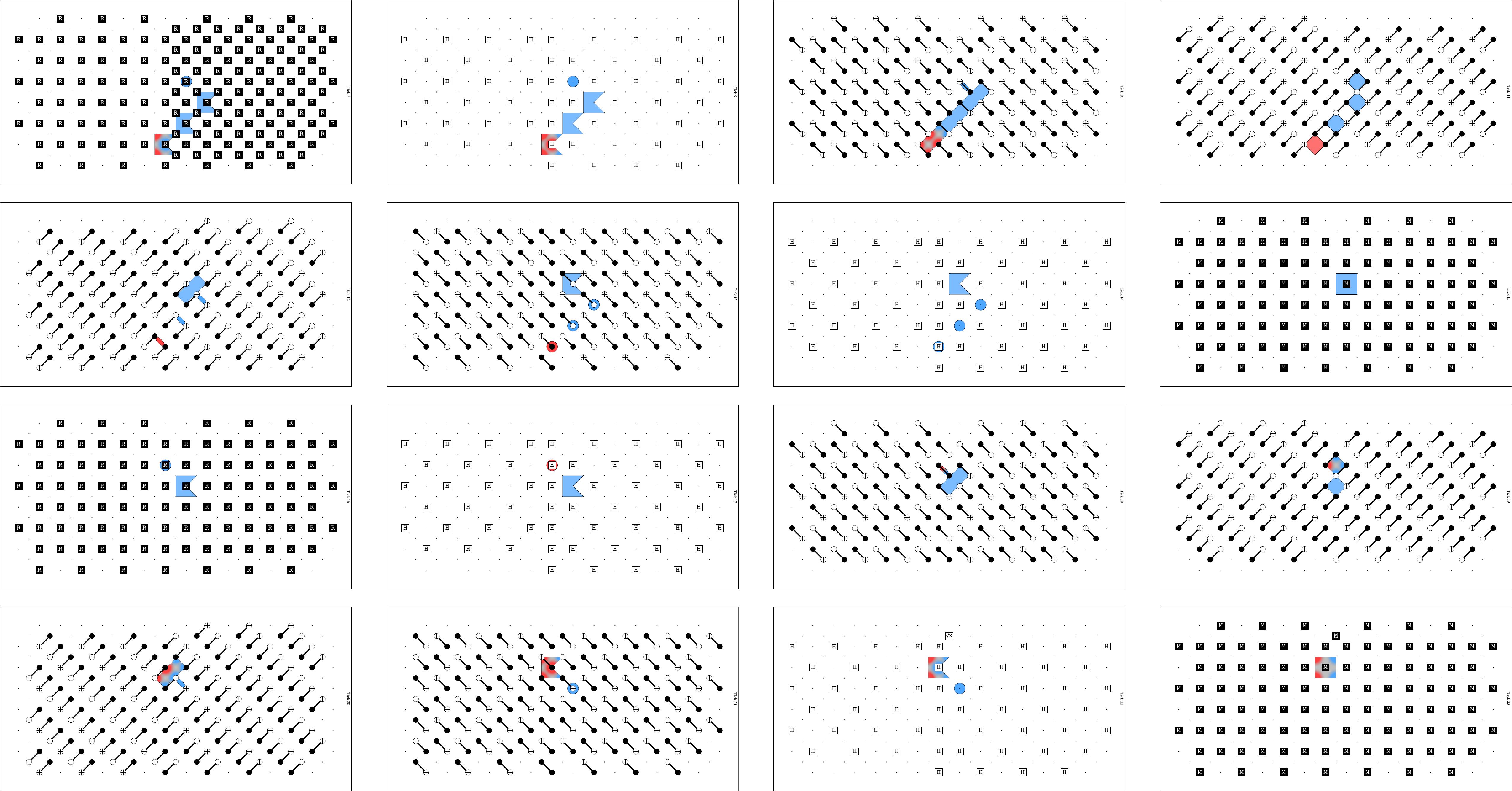}
    \caption{ The set of detectors triggered by each error in the shortest error chain that limits the fault distance to $d-1$ for the proposed (NL) method.}
    \label{fig:circuit_detslice_proposed_non_local_hook_error}
\end{figure*}

\begin{figure*}[h]
    \centering
    \includegraphics[scale=0.1]{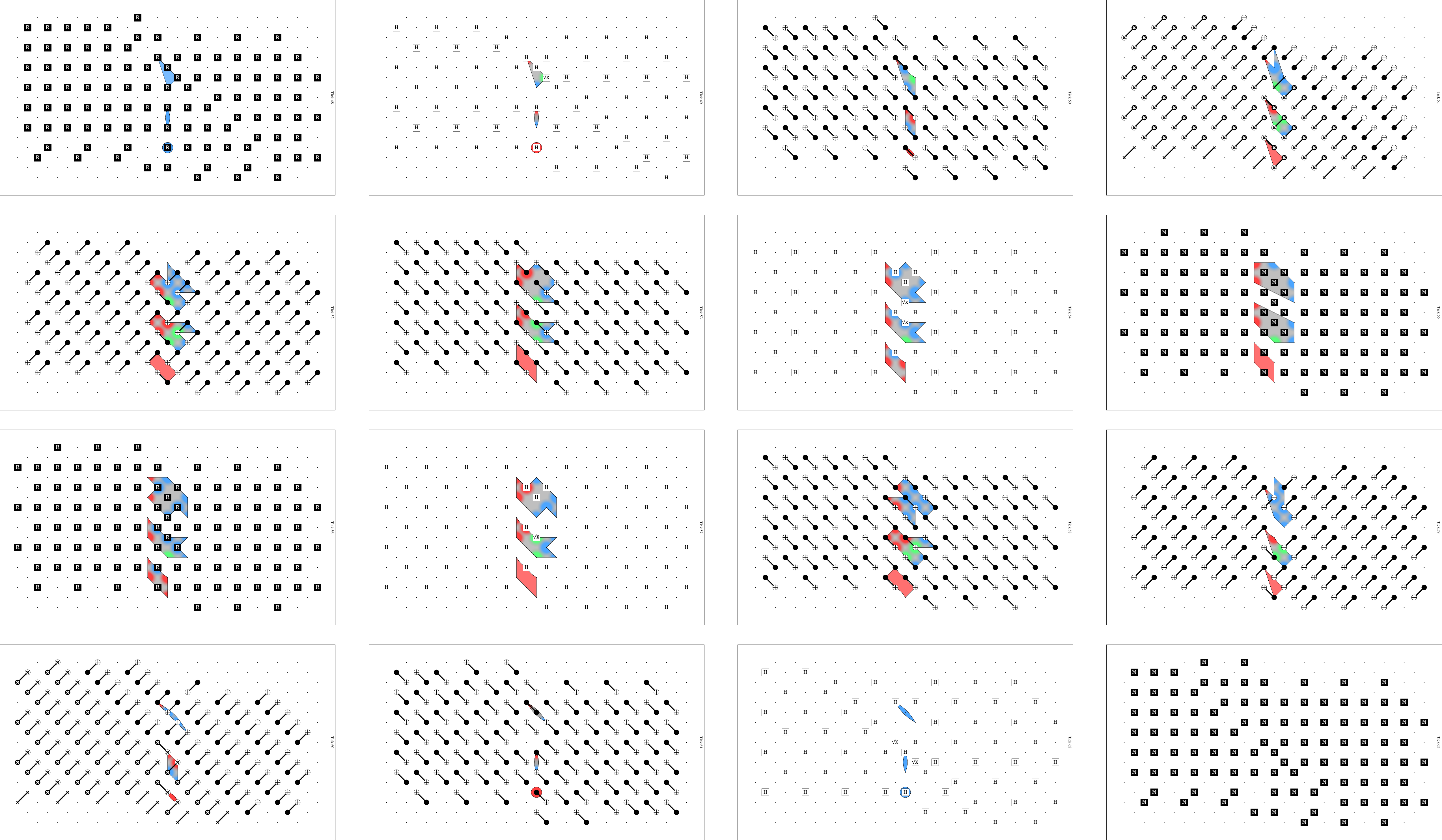}
    \caption{ The set of detectors triggered by each error in the shortest error chain that limits the fault distance to $d-3$ for the proposed (L) method.}
    \label{fig:circuit_detslice_proposed_local_hook_error}
\end{figure*}

The results are shown in Tab.~\ref{tab:fault_distance}. From the results, Bombín's method halved the fault distance, and the proposed method of NL and L reduced the fault distances by one and three, respectively. The error chains for each method are shown Fig.~\ref{fig:circuit_detslice_bombin_hook_error}, \ref{fig:circuit_detslice_proposed_non_local_hook_error} and \ref{fig:circuit_detslice_proposed_local_hook_error}. These figures highlight the detectors triggered by each error in the shortest error chain. In Bombín's method, as mentioned earlier, hook errors occur when the twist defect is braided through the bulk of the surface code. Since the CNOT gates in the syndrome measurement for either Z or X stabilizers follow an N-shaped ordering, these hook errors cannot be avoided, thereby reducing the fault distance.
 Similarly, the hook errors in both the proposed (NL) and (L) methods correspond to constant-length error chains. Consequently, even as the code distance increases, the upper bounds of the fault distances remain capped at $d-1$ and $d-3$, respectively. These results indicate that hook errors arising from twist defect braiding within the bulk of the surface code are unavoidable with the current scheduling.

\subsection{Logical Error Rates}

\begin{figure*}[b]
    \centering
    \includegraphics[scale=0.70]{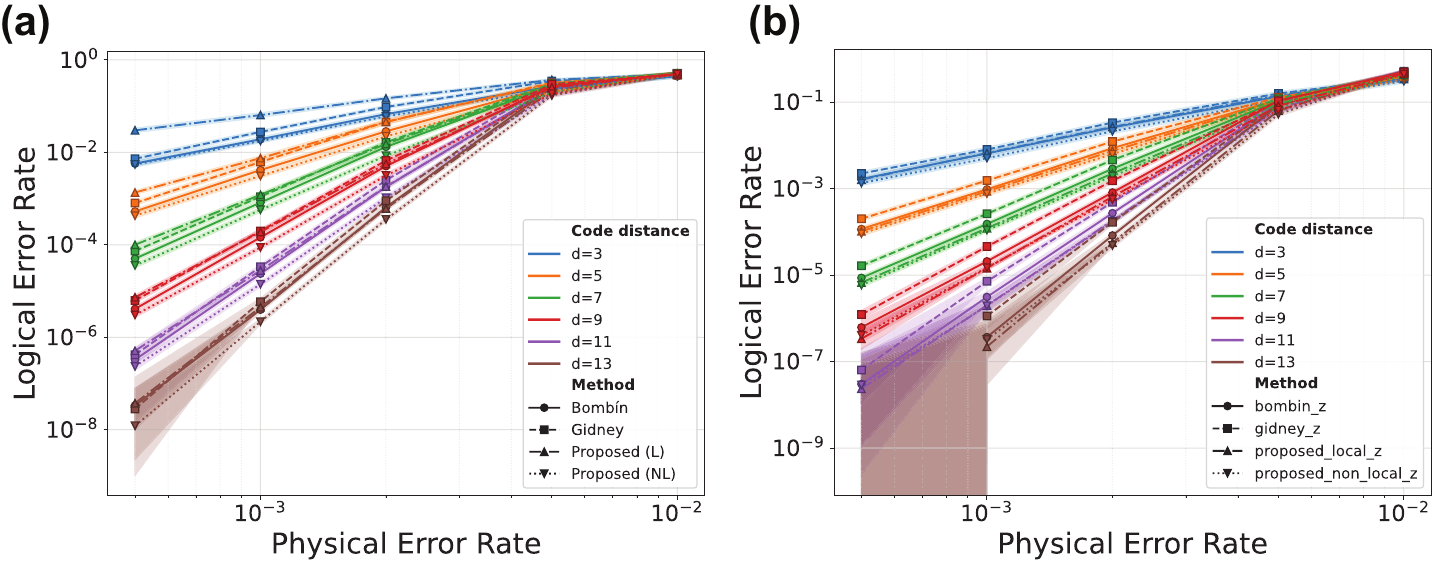}
    \caption{Numerical results of logical error rates under circuit-level noise. The logical error rates are shown for the logical operator transformations (a)~X$\rightarrow$Y and (b)~Z$\rightarrow$Z.}
    \label{fig:comparison_of_all_methods}
\end{figure*}

\begin{figure*}[h]
    \centering
    \includegraphics[scale=0.43]{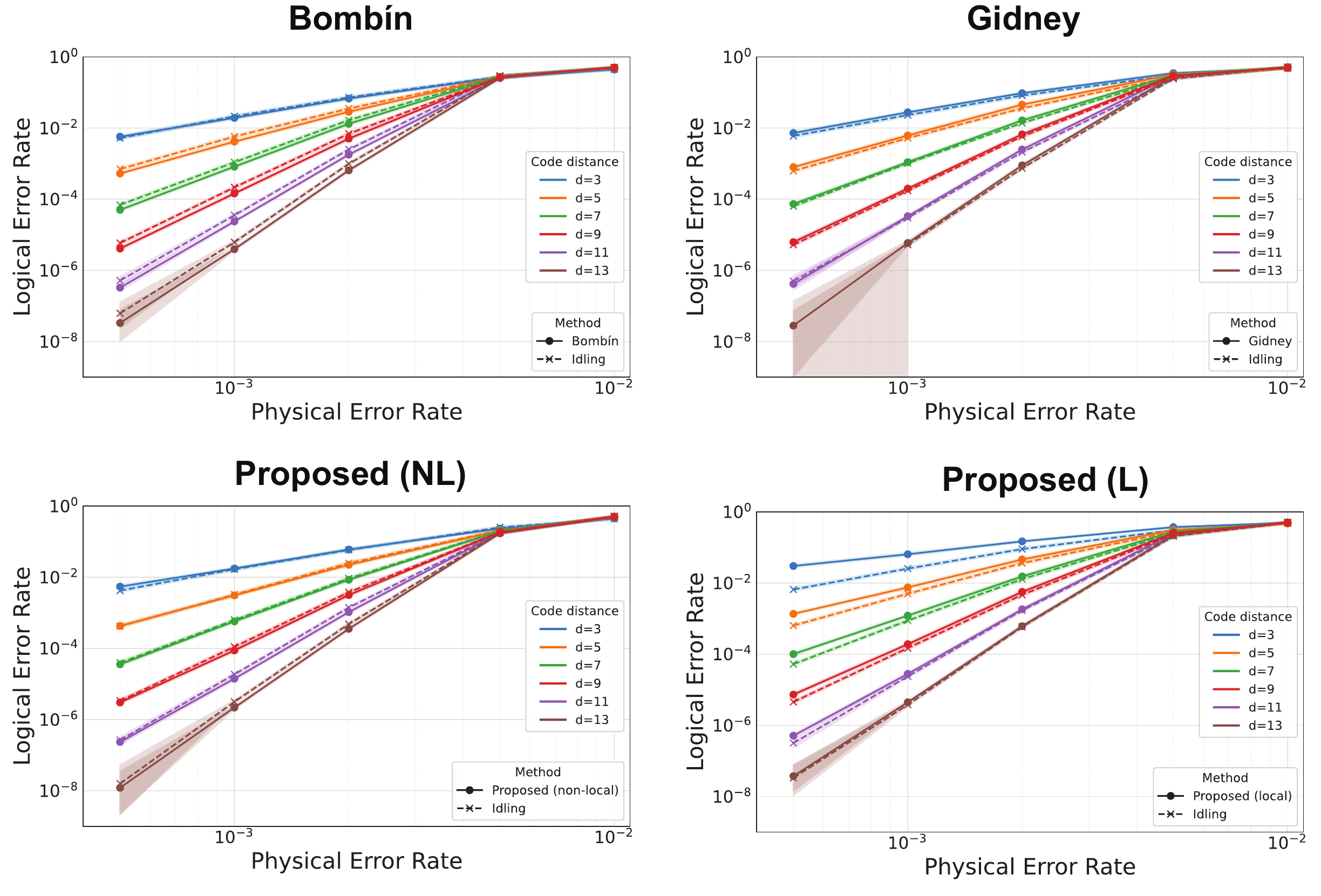}
    \caption{Comparison of all methods and the idling circuits that have the same spacetime volumes as each method. The circuits were initialized in the X basis.}
    \label{fig:comparison_with_idling}
\end{figure*}

Second, we calculated the logical error rates for all methods and the idling circuits that have volumes equivalent to each method using Monte Carlo sampling. We employed PyMatching~\cite{Higgott2025sparseblossom} as a decoder, with hyperedges in the detector error models decomposed into graph-like edges using Stim. The numerical results of the comparison of all methods are shown in Fig.~\ref{fig:comparison_of_all_methods}. These results indicate that, despite slight reductions in fault distances, the logical error rates remain comparable across all methods for large code distances ($\ge 5$). This suggests that the error events reducing the fault distance are significantly rarer than those causing length-$d$ error chains. The numerical results comparing all methods with the corresponding idling circuits are shown in Fig.~\ref{fig:comparison_with_idling}. Each idling circuit was designed to have a spacetime volume equivalent to that of the corresponding method and was initialized in the X basis. The results demonstrate that the logical error rates of all methods are comparable to those of their respective idling circuits, indicating that the methods effectively suppress errors. It is worth noting that for Bombín's method, the logical error rate is slightly lower than that of the idling circuit. This is because the idling circuit, which idles for the same duration as the SWAP operations in Bombín's method, was not subjected to the shrinking operations (via SWAP gates) used in Bombín's method, which effectively reduced the active volume in the latter.

\section{Conclusion}\label{sec:conclusion}
In this work, we provided circuit-level implementations for existing methods that perform a logical S gate in the surface code, and proposed a new method that implements the gate with less overhead than all existing methods, namely \(2d \times d \times d\). We then outlined a design for braiding a twist defect in the bulk of the surface code, which makes it possible to implement the proposed method. We presented two circuit-level implementations of the proposed method: one requires some constant-length non-local two-qubit gates, and the other requires only local gates, including CXSWAP gates. We considered that the non-local implementation is sufficient for devices that allow non-local gates, such as neutral-atom systems. However, to implement the proposed method on devices where non-local gates are challenging, a local implementation is necessary. We evaluate all methods in this work using Stim. From graph-like and heuristic searches, the proposed method exhibits a constant reduction in fault distance by one and three for the non-local and local variants, respectively. Furthermore, we calculated the logical error rates for all methods, showing that the proposed methods achieve comparable performance in the large code distance (\(\ge 5\)) regime while reducing spacetime volume. We believe that the proposed method is the most promising for near-term quantum computers.

Several directions for future work remain. First, since the proposed methods in this work do not achieve the full fault distance, they may become less effective if physical error rates become sufficiently small. Therefore, future work should focus on improving the fault distance. Second, we also aim to reduce the resources required for the logical H gate by employing the twist defect braiding techniques explored in this study. The volume of the logical H gate is typically \(2d \times 3d \times d\) \cite{litinski2019game, geher2024error}, making it more costly than the logical S gate, despite H gates being also frequently used in quantum circuits. Third, in this work, we propose a method to braid a twist defect in the bulk of the surface code. Thus, it might make it possible to realize logical operations that utilize twist-defect braiding in the surface code \cite{bombin2010topological,bombin2023logical,chamberland2022circuit,hastings2014reduced}.

\section*{Acknowledgments}
YH proposed the original idea, conducted the numerical calculations, and prepared the manuscript. 
SI contributed to discussions on the optimization of circuits. 
YU and YS supervised the project. 
All authors reviewed, revised, and approved the final manuscript. 
This work is supported by MEXT Q-LEAP Grant No.~JPMXS0120319794 and JPMXS0118068682, MEXT Feasibility Study on the future HPCI, JST Moonshot R\&D Grant No.~JPMJMS2061, JST CREST Grant No.~JPMJCR23I4, JPMJCR24I4, and JPMJCR25I4, JSPS KAKENHI Grant No.~JP22H05000 and JP25K21176.
\clearpage

\bibliography{ref}

@article{brown2017poking,
  title={Poking holes and cutting corners to achieve Clifford gates with the surface code},
  author={Brown, Benjamin J and Laubscher, Katharina and Kesselring, Markus S and Wootton, James R},
  journal={Physical Review X},
  volume={7},
  number={2},
  pages={021029},
  year={2017},
  publisher={APS}
}

@article{bombin2023logical,
  title={Logical blocks for fault-tolerant topological quantum computation},
  author={Bombin, Hector and Dawson, Chris and Mishmash, Ryan V and Nickerson, Naomi and Pastawski, Fernando and Roberts, Sam},
  journal={PRX Quantum},
  volume={4},
  number={2},
  pages={020303},
  year={2023},
  publisher={APS}
}

@article{gidney2024inplace,
  title={Inplace access to the surface code y basis},
  author={Gidney, Craig},
  journal={Quantum},
  volume={8},
  pages={1310},
  year={2024},
  publisher={Verein zur F{\"o}rderung des Open Access Publizierens in den Quantenwissenschaften}
}

@article{moussa2016transversal,
  title={Transversal Clifford gates on folded surface codes},
  author={Moussa, Jonathan E},
  journal={Physical Review A},
  volume={94},
  number={4},
  pages={042316},
  year={2016},
  publisher={APS}
}

@article{mcewen2023relaxing,
  title={Relaxing hardware requirements for surface code circuits using time-dynamics},
  author={McEwen, Matt and Bacon, Dave and Gidney, Craig},
  journal={Quantum},
  volume={7},
  pages={1172},
  year={2023},
  publisher={Verein zur F{\"o}rderung des Open Access Publizierens in den Quantenwissenschaften}
}

@article{gidney2021stim,
  doi = {10.22331/q-2021-07-06-497},
  url = {https://doi.org/10.22331/q-2021-07-06-497},
  title = {Stim: a fast stabilizer circuit simulator},
  author = {Gidney, Craig},
  journal = {{Quantum}},
  issn = {2521-327X},
  publisher = {{Verein zur F{\"{o}}rderung des Open Access Publizierens
                in den Quantenwissenschaften}},
  volume = 5,
  pages = 497,
  month = jul,
  year = 2021
}

@article{lacroix2025scaling,
  title={Scaling and logic in the color code on a superconducting quantum processor},
  author={Lacroix, Nathan and Bourassa, Alexandre and Heras, Francisco JH and Zhang, Lei M and Bausch, Johannes and Senior, Andrew W and Edlich, Thomas and Shutty, Noah and Sivak, Volodymyr and Bengtsson, Andreas and others},
  journal={Nature},
  pages={1--3},
  year={2025},
  publisher={Nature Publishing Group UK London}
}

@article{google2025quantum,
  title={Quantum error correction below the surface code threshold},
  author={Google Quantum AI and Collaborators},
  journal={Nature},
  volume={638},
  number={8052},
  pages={920--926},
  year={2025},
  publisher={Nature Publishing Group UK London}
}

@article{rosenfeld2025magic,
  title={Magic state cultivation on a superconducting quantum processor},
  author={Rosenfeld, Emma and Gidney, Craig and Roberts, Gabrielle and Morvan, Alexis and Lacroix, Nathan and Kafri, Dvir and Marshall, Jeffrey and Li, Ming and Sivak, Volodymyr and Abanin, Dmitry and others},
  journal={arXiv preprint arXiv:2512.13908},
  year={2025}
}

@article{Higgott2025sparseblossom,
  doi = {10.22331/q-2025-01-20-1600},
  url = {https://doi.org/10.22331/q-2025-01-20-1600},
  title = {Sparse {B}lossom: correcting a million errors per core second with minimum-weight matching},
  author = {Higgott, Oscar and Gidney, Craig},
  journal = {{Quantum}},
  issn = {2521-327X},
  publisher = {{Verein zur F{\"{o}}rderung des Open Access Publizierens in den Quantenwissenschaften}},
  volume = {9},
  pages = {1600},
  month = jan,
  year = {2025}
}

@inproceedings{maurya2025synchronization,
  title={Synchronization for Fault-Tolerant Quantum Computers},
  author={Maurya, Satvik and Tannu, Swamit},
  booktitle={Proceedings of the 52nd Annual International Symposium on Computer Architecture},
  pages={1370--1385},
  year={2025}
}

@article{gidney2024magic,
  title={Magic state cultivation: growing T states as cheap as CNOT gates},
  author={Gidney, Craig and Shutty, Noah and Jones, Cody},
  journal={arXiv preprint arXiv:2409.17595},
  year={2024}
}

@article{geher2024error,
  title={Error-corrected Hadamard gate simulated at the circuit level},
  author={Geh{\'e}r, Gy{\"o}rgy P and McLauchlan, Campbell and Campbell, Earl T and Moylett, Alexandra E and Crawford, Ophelia},
  journal={Quantum},
  volume={8},
  pages={1394},
  year={2024},
  publisher={Verein zur F{\"o}rderung des Open Access Publizierens in den Quantenwissenschaften}
}

@article{horsman2012surface,
  title={Surface code quantum computing by lattice surgery},
  author={Horsman, Dominic and Fowler, Austin G and Devitt, Simon and Van Meter, Rodney},
  journal={New Journal of Physics},
  volume={14},
  number={12},
  pages={123011},
  year={2012},
  publisher={IOP Publishing}
}

@article{bombin2009quantum,
  title={Quantum measurements and gates by code deformation},
  author={Bomb{\'\i}n, H{\'e}ctor and Martin-Delgado, Miguel Angel},
  journal={Journal of Physics A: Mathematical and Theoretical},
  volume={42},
  number={9},
  pages={095302},
  year={2009},
  publisher={IOP Publishing}
}

@article{litinski2019game,
  title={A game of surface codes: Large-scale quantum computing with lattice surgery},
  author={Litinski, Daniel},
  journal={Quantum},
  volume={3},
  pages={128},
  year={2019},
  publisher={Verein zur F{\"o}rderung des Open Access Publizierens in den Quantenwissenschaften}
}

@article{higgott2022pymatching,
  title={Pymatching: A python package for decoding quantum codes with minimum-weight perfect matching},
  author={Higgott, Oscar},
  journal={ACM Transactions on Quantum Computing},
  volume={3},
  number={3},
  pages={1--16},
  year={2022},
  publisher={ACM New York, NY}
}

@article{zhang2025demonstrating,
  title={Demonstrating a universal logical gate set in error-detecting surface codes on a superconducting quantum processor},
  author={Zhang, Jiaxuan and Chen, Zhao-Yun and Wang, Yun-Jie and Lu, Bin-Han and Zhang, Hai-Feng and Li, Jia-Ning and Duan, Peng and Wu, Yu-Chun and Guo, Guo-Ping},
  journal={npj Quantum Information},
  volume={11},
  number={1},
  pages={177},
  year={2025},
  publisher={Nature Publishing Group UK London}
}

@article{sahay2025error,
  title={Error correction of transversal cnot gates for scalable surface-code computation},
  author={Sahay, Kaavya and Lin, Yingjia and Huang, Shilin and Brown, Kenneth R and Puri, Shruti},
  journal={PRX quantum},
  volume={6},
  number={2},
  pages={020326},
  year={2025},
  publisher={APS}
}

@article{yoder2017surface,
  title={The surface code with a twist},
  author={Yoder, Theodore J and Kim, Isaac H},
  journal={Quantum},
  volume={1},
  pages={2},
  year={2017},
  publisher={Verein zur F{\"o}rderung des Open Access Publizierens in den Quantenwissenschaften}
}

@article{chen2024transversal,
  title={Transversal Logical Clifford gates on rotated surface codes with reconfigurable neutral atom arrays},
  author={Chen, Zi-Han and Chen, Ming-Cheng and Lu, Chao-Yang and Pan, Jian-Wei},
  journal={arXiv preprint arXiv:2412.01391},
  year={2024}
}

@article{hirano2025efficient,
  title={Efficient magic state cultivation with lattice surgery},
  author={Hirano, Yutaka and Toshio, Riki and Itogawa, Tomohiro and Fujii, Keisuke},
  journal={arXiv preprint arXiv:2510.24615},
  year={2025}
}

@article{vaknin2025efficient,
  title={Efficient magic state cultivation on the surface code},
  author={Vaknin, Yotam and Jacoby, Shoham and Grimsmo, Arne and Retzker, Alex},
  journal={arXiv preprint arXiv:2502.01743},
  year={2025}
}

@article{litinski2018lattice,
  title={Lattice surgery with a twist: Simplifying Clifford gates of surface codes},
  author={Litinski, Daniel and von Oppen, Felix},
  journal={Quantum},
  volume={2},
  pages={62},
  year={2018},
  publisher={Verein zur F{\"o}rderung des Open Access Publizierens in den Quantenwissenschaften}
}

@article{hirai2026no,
  title={No More Hooks in the Surface Code: Distance-Preserving Syndrome Extraction for Arbitrary Layouts at Minimum Depth},
  author={Hirai, Yuga and Ikari, Shota and Ueno, Yosuke and Suzuki, Yasunari},
  journal={arXiv preprint arXiv:2603.01628},
  year={2026}
}

@misc{gidney_crumble,
  title        = {Crumble, a prototype stabilizer circuit editor},
  auther = {Craig Gidney},
  howpublished = {\url{https://algassert.com/crumble}},
  year ={2026}
}

@article{hastings2014reduced,
  title={Reduced space-time and time costs using dislocation codes and arbitrary ancillas},
  author={Hastings, Matthew B and Geller, Alan},
  journal={arXiv preprint arXiv:1408.3379},
  year={2014}
}

@article{bombin2010topological,
  title={Topological order with a twist: Ising anyons from an abelian model},
  author={Bomb{\'\i}n, H{\'e}ctor},
  journal={Physical review letters},
  volume={105},
  number={3},
  pages={030403},
  year={2010},
  publisher={APS}
}

@article{chamberland2022circuit,
  title={Circuit-level protocol and analysis for twist-based lattice surgery},
  author={Chamberland, Christopher and Campbell, Earl T},
  journal={Physical Review Research},
  volume={4},
  number={2},
  pages={023090},
  year={2022},
  publisher={APS}
}

@misc{stimcircuit_github,
  title        = {All circuits in Stim are available here},
  howpublished = {\url{https://github.com/yugahirai/logical_s_gate}},
  year ={2026}
}

@article{stephens2014fault,
  title={Fault-tolerant thresholds for quantum error correction with the surface code},
  author={Stephens, Ashley M},
  journal={Physical Review A},
  volume={89},
  number={2},
  pages={022321},
  year={2014},
  publisher={APS}
}

@article{fowler2009high,
  title={High-threshold universal quantum computation on the surface code},
  author={Fowler, Austin G and Stephens, Ashley M and Groszkowski, Peter},
  journal={Physical Review A—Atomic, Molecular, and Optical Physics},
  volume={80},
  number={5},
  pages={052312},
  year={2009},
  publisher={APS}
}
\end{document}